\documentclass{aa} 

\usepackage{graphicx}
\usepackage{txfonts}
\usepackage{natbib}
\bibpunct{(}{)}{;}{a}{}{,}

\newcommand{\be}{\begin{equation}}
\newcommand{\ee}{\end{equation}}
\newcommand{\angstrom}{\mbox{\normalfont\AA}}

\def\apjl{ApJL}
\def\apjs{ApJS}

\newcommand{\Mpc}{$h^{-1}$\thinspace Mpc}

\newcommand{\vmh}{h^{-1}\mathrm{Mpc} }

\begin{document}  

\title{
Galaxy groups and clusters and their brightest galaxies within the cosmic web
}

\author {Maret~Einasto\inst{1} 
\and Jaan~Einasto\inst{1,2,3}
\and Peeter~Tenjes\inst{1} 
\and Suvi~Korhonen\inst{4}
\and Rain~Kipper\inst{1} 
\and Elmo~Tempel\inst{1,2} 
\and Lauri Juhan~Liivam\"agi\inst{1} 
\and Pekka~Hein\"am\"aki\inst{4}
}
\institute{Tartu Observatory, University of Tartu, Observatooriumi 1, 61602 T\~oravere, Estonia
\and
Estonian Academy of Sciences, Kohtu 6, 10130 Tallinn, Estonia
\and
ICRANet, Piazza della Repubblica 10, 65122 Pescara, Italy
\and 
Tuorla Observatory, Department of Physics and Astronomy, University of Turku, Vesilinnantie 5, 20014
Turku, Finland
}

\authorrunning{Einasto, M. et al. }

\offprints{Einasto, M.}

\date{ Received   / Accepted   }

\titlerunning{Galaxy groups in the cosmic web}

\abstract
{
The evolution of galaxy groups and   the brightest group galaxies (BGGs)
is influenced by their location in the cosmic web. 
}
{
Our aim is to combine data on galaxy groups, their BGGs, 
and their location in the cosmic web, to  determine classes of groups and clusters, and to 
obtain a better understanding of their properties and evolution.
}
{
Data on groups and their BGGs
are based on the Sloan Digital Sky Survey DR10 MAIN spectroscopic galaxy sample in the redshift 
range $0.009 \leq z \leq 0.200$. We characterize the group environments by the luminosity--density
field and their filament membership. We  divide BGGs according to their star formation properties as quenched (Q), red
star-forming galaxies (RSF), and blue star-forming galaxies (BSF). 
We apply multidimensional Gaussian mixture  modelling
to divide groups based on the properties of 
the groups, their BGGs, and their environments. We analyse the offset of BGGs with respect
to the group centre, and the relation between the stellar velocity dispersion of BGGs $\sigma^\star$
and the group velocity dispersions $\sigma_{\mathrm{v}}$. 
For comparison we also analyse the properties of single galaxies of different star formation properties
in various environments.
}
{
The galaxy groups in our sample can be divided  into two main classes: high-luminosity rich groups
and clusters, and low-luminosity poor groups
with threshold luminosity  $L^{thr}_{gr} = 15 \times10^{10} h^{-2} L_{\sun}$
and total mass $M^{thr}_{gr} \approx 23 \times10^{12} h^{-1} M_{\sun}$. 
The brightest galaxies in clusters and groups have different star formation properties.
In rich groups and clusters  $\approx 90$~\% of the  BGGs 
are red quenched galaxies,  while in poor groups only $\approx 40- 60$~\% of BGGs
are red and quenched, and the rest of the BGGs are star-forming, either blue 
($20 - 40$~\% of BGGs) or red  ($\sim$17\% of BCGs).  
Rich groups and clusters are located in global high-density regions 
 (superclusters) 
in filaments or filament outskirts, while 
poor groups reside everywhere in the cosmic web regardless of the global density (superclusters or voids).
Clusters with quenched BGGs  have higher luminosities and their BGGs are closer to the cluster centre than
in clusters with star-forming BGGs. Groups of the same richness with red (quenched and star-forming) BGGs
are more luminous, and they lie in higher global density environment than groups with
blue star-forming BGGs.
}
{
Our results suggest that the evolution of  groups and clusters and their BGGs 
is related to their location  in the cosmic web. We emphasize the role of  
global high-density regions--superclusters as a special environment
for group growth. The processes that shape the properties of 
groups and their BGG are different and/or have 
different timescales in groups and clusters. 
}


\maketitle

\section{Introduction} 
\label{sect:intro} 

Most galaxies in the cosmic web lie in groups of various richness from galaxy pairs 
to the richest clusters.
Simulations show that galaxies and their systems (groups and clusters) 
can form in the cosmic density field where large-scale density perturbations 
in combination with small-scale overdensity are sufficiently high 
\citep{2011A&A...534A.128E, 2011A&A...531A.149S, 2021arXiv210602672P}. 
Structure formation is modulated by the combination of density waves: 
voids form where negative phases of density waves combine, 
and rich clusters and superclusters form where positive phases combine. 
As a result, the richest, often X-ray, merging clusters typically
lie in the highest density regions of 
the cosmic web, which are the deepest potential wells, in superclusters and their high-density cores
\citep{1999ApJ...521...90H, 2005A&A...444..387D, 2005AdSpR..36..630B, 2016A&A...595A..70E,
2017ApJ...844...25B, 2021A&A...649A..51E}.
Low-density regions of the cosmic web are populated by 
poor groups and galaxies, which may not belong to any detectable groups
\citep[][and references therein]{2011A&A...534A.128E,2012A&A...545A.104L, 2022A&A...668A..69E, 2023arXiv230414387J}.
Richer galaxy groups tend to have higher  connectivity than poor groups; 
groups in superclusters also have higher connectivity than
groups of the same richness in voids \citep[][and references therein]{2019MNRAS.489.5695D, 2020A&A...641A.172E, 
2021A&A...649A..51E}. 

The brightest galaxies in clusters are the most luminous and massive galaxies in the Universe,
often located near the centres of cluster potential wells.
Early studies of the cosmic web have already shown that the brightest galaxies of clusters 
in superclusters are elongated along the supercluster axis, suggesting that the evolution of 
clusters and their brightest galaxies is related to the environment where they
reside \citep{1978MNRAS.185..357J}. 
Observations and simulations show that the BGGs of rich galaxy clusters are more luminous and have higher
stellar masses and larger stellar velocity dispersions than BGGs of poor groups
\citep[][and references therein]{1978MNRAS.185..357J, 2020ApJ...891..129S, 2021A&A...649A..42C, 
2021MNRAS.507.5780M, 2022A&A...668A..69E}.
The stellar velocity dispersion of BGGs, $\sigma^\star$,
is proportional to the group velocity dispersion, $\sigma_{\mathrm{v}}$
\citep[][and references therein]{2000ASPC..209..360E, 2020ApJ...891..129S, 2021MNRAS.507.5780M}.

There is no clear distinction between groups and clusters. \citet{2015AJ....149...54T}
emphasized that there does not seem to be a meaningful (or useful) threshold between them.
However, recently \citet{2022A&A...668A..69E} found a clear limit of group luminosity between 
groups in the lowest global density regions and at higher density:
high-luminosity groups with luminosity 
$L_{gr} \geq 15\times10^{10} h^{-2} L_{\sun}$ are absent from the global lowest density
regions. \citet{2022A&A...668A..69E} divided groups into high- and low-luminosity classes
at this luminosity limit. 
They also found that while the brightest group galaxies (BGGs) of high-luminosity groups 
are almost all quenched galaxies with old stellar populations,
a large percentage of BGGs of low-luminosity groups are still forming stars.

In this study our aim is to search for the division of groups and clusters 
 using information on the properties of groups, on the star formation properties of their BGGs,  
 and on group environment. 
We divide the   BGGs of groups  into quenched galaxies with no active star formation, red 
star-forming galaxies, and blue star-forming galaxies, and study whether 
groups of different luminosity have similar BGGs.
The environment of groups is defined in  two different ways. First, we  use
data on the global environment of groups, quantified using the luminosity--density field.
Using certain threshold density limits, we divide the cosmic web according to these limits
as global high-density regions or superclusters, and global low-density regions or voids.
We  compare the groups in superclusters with those in voids to
learn about whether  group properties differ in various environments. 
We would like to understand whether superclusters and their
high-density cores form a special environment for group formation and growth, 
and whether  poor groups in superclusters are
different from
those in low-density environments between superclusters.

Second, we analyse information on the filament membership of groups. 
Our analysis brings us to the questions of whether groups of different luminosities
and BGG properties are connected
to filaments in a similar or a different way, and whether groups and/or their BGGs in or near filaments
differ from groups far from filaments.
We may speculate that luminous high-connectivity groups  in global high-density regions
also have more mature BGGs.
Therefore, one focus of our study is a comparison of groups with quenched galaxies, and red and blue star-forming
BGGs and their environments.

Among the dynamical properties of groups and clusters we analyse the location of BGGs with respect
to the group centre, and compare the relation between the stellar velocity dispersion of BGGs ($\sigma^\star$)
and the group velocity dispersions ($\sigma_{\mathrm{v}}$) for groups with different BGGs.
In this way we aim to obtain a better understanding of the properties and coevolution of
groups and their BGGs in the cosmic web.

Our study is based on the Sloan Digital Sky Survey (SDSS) DR10 MAIN spectroscopic galaxy sample in the redshift 
range $0.009 \leq z \leq 0.200$.
We used this sample to calculate the luminosity--density
field of galaxies, to determine groups and filaments in the galaxy distribution,
and to obtain data on galaxy properties
\citep{2011ApJS..193...29A, 2014ApJS..211...17A}.
The luminosity--density field with smoothing length $8$~\Mpc, $D8$,
characterizes the global environment of galaxies.
To include data on faint groups, we chose the sample of groups  from the 
redshift range $0.03 \leq z \leq 0.08$. 
We note that \citet{2022A&A...668A..69E} used a higher redshift range,  $0.07 \leq z \leq 0.10$,
in order to include rich superclusters at redshifts $z \approx 0.08 - 0.1$
(the Sloan Great Wall and other superclusters). Our present sample covers 
the region of the Hercules supercluster at the   redshift range $z \approx 0.03 - 0.04$ 
and an underdense region between the Hercules supercluster and the Corona Borealis and Bootes
superclusers at redshift $z \approx 0.07$. 
In order to understand better the properties of BGGs of the poorest groups, we
also included in our study  single galaxies. Single galaxies are galaxies that do not
belong to any detectable groups within the SDSS spectroscopic sample luminosity limits.

As in  \citet{2022A&A...668A..69E}, we applied 
the following cosmological parameters: the Hubble parameter $H_0=100~ 
h$ km~s$^{-1}$ Mpc$^{-1}$, matter density $\Omega_{\rm m} = 0.27$, 
parameter $h = 0.7$,  and 
dark energy density $\Omega_{\Lambda} = 0.73$ 
\citep{2011ApJS..192...18K}.

\section{Observational data} 
\label{sect:galdat} 

Our study is based on data from the SDSS DR10 MAIN spectroscopic galaxy sample,
which is used to compile catalogues of  galaxy groups and filaments and to calculate
the luminosity--density field.
Galaxies in this sample have
apparent Galactic extinction-corrected $r$ magnitudes $r \leq 
17.77$ and redshifts $0.009 \leq z \leq 0.200$
\citep{2011ApJS..193...29A, 2014ApJS..211...17A}.
The absolute magnitudes of galaxies are calculated as  
\begin{equation}
M_r = m_r - 25 -5\log_{10}(d_L)-K,
\end{equation} 
where $d_L$ is the luminosity distance in units of $h^{-1}$Mpc. Here
$K$ is the $k$+$e$-correction calculated as in 
\citet{2007AJ....133..734B} and  \citet{2003ApJ...592..819B}. For a detailed description we refer 
to \citet{2014A&A...566A...1T}.

{\it Galaxy groups, their brightest galaxies, and single galaxies}.
Based on the SDSS MAIN galaxy sample, 
\citet{2014A&A...566A...1T} generated a catalogue
of galaxy groups, applying the   friends-of-friends (FoF) clustering analysis
method \citep{1982Natur.300..407Z, 1982ApJ...257..423H}. 
This method works as follows: in the neighbourhood of every galaxy up to a certain linking 
length, neighbouring galaxies are searched for. Every galaxy closer than a linking length to any member of
a group is considered a  member of a group. 

In a flux-limited sample, such  as the SDSS MAIN galaxy sample, the density of galaxies slowly 
decreases with distance. To take this selection effect into account, the 
linking length is re-scaled with distance, so that the scaling relation
was calibrated using observed groups. As a result, the 
maximum sizes in the sky projection and the velocity dispersions of the groups 
are similar at all distances. The  redshift-space distortions (also known as the  Fingers of God) 
for groups were suppressed,
as described in detail in \citet{2014A&A...566A...1T}.
The luminosities of groups $L_{gr}$ in this catalogue 
are calculated
using the $r$-band luminosity of group member galaxies.

\begin{figure}
\centering
\resizebox{0.46\textwidth}{!}{\includegraphics[angle=0]{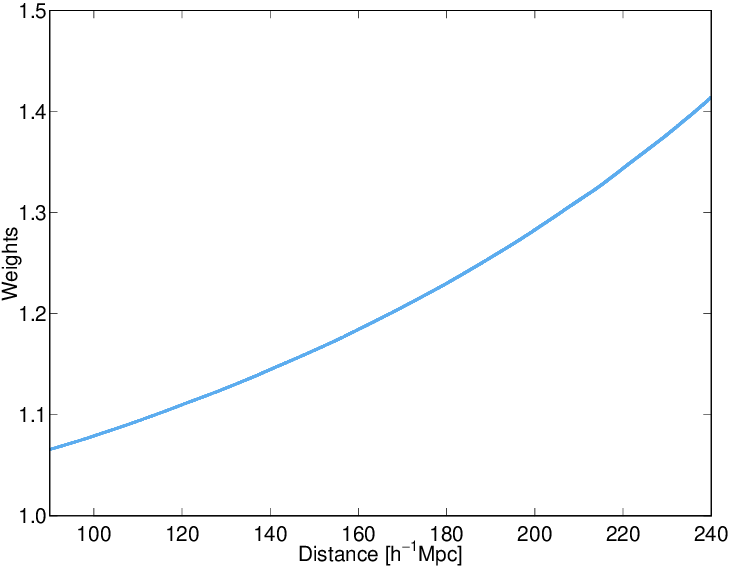}}
\caption{Weights used to correct for unobserved group members outside
the observational luminosity window vs distance of groups.
}
\label{fig:wdist}
\end{figure}

\begin{figure}
\centering
\resizebox{0.46\textwidth}{!}{\includegraphics[angle=0]{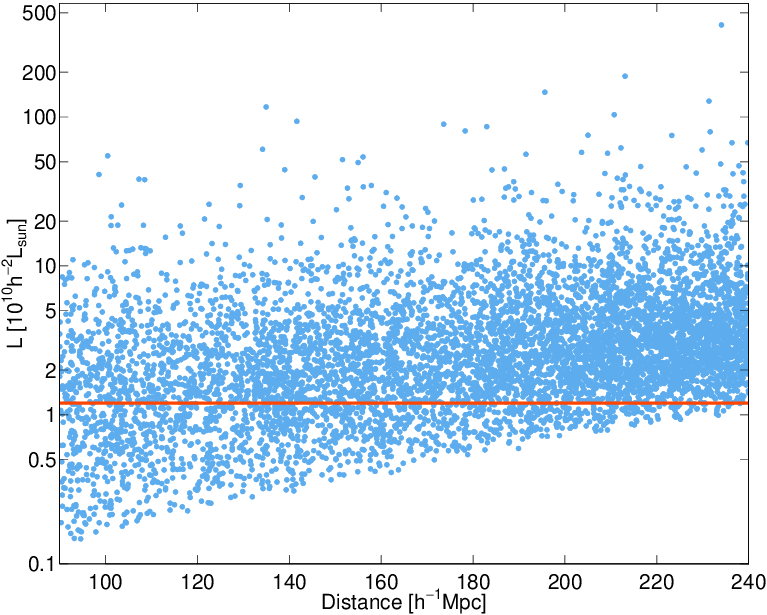}}
\caption{Luminosity of groups vs distance for
redshift range $0.03 \leq z \leq 0.08$. The red line shows the adopted luminosity limit,
$L_{gr} = 1.2 \times10^{10} h^{-2} L_{\sun}$,
for a complete sample of groups. To make the figure file smaller,   only  
one-quarter of all the groups, randomly chosen, are plotted. 
}
\label{fig:lgrdist}
\end{figure}

Groups may also contain galaxies that lie outside the observational window. To take this
effect into account, we corrected luminosities of groups for the missing (unobserved)
galaxies at a distance of a given group, so that the estimated total luminosity of groups,
which also takes into account   the expected luminosities of the unobserved galaxies, is
\begin{equation}
L_\mathrm{tot}=L_\mathrm{obs}\cdot W_d,
\label{eq:totlum}
\end{equation}
where $L_\mathrm{obs}$ is the observed luminosity of the galaxy. 
The luminosity weights versus distance of a group are plotted in Fig.~\ref{fig:wdist}.
The value of weights increases with distance as more galaxies remain outside of the observational window.
Figure~\ref{fig:lgrdist} presents luminosity of groups versus their distance.
To have a complete sample of groups, which also includes   faint groups, we applied
distance limits $90 - 240$~\Mpc\ (redshift range $0.03 \leq z \leq 0.08$). As the SDSS sample cone that is  
close to us is  very narrow, we did not use data on very close galaxies and groups.
In Fig.~\ref{fig:wdist}  we see that within our chosen distance limits the  weights are  slightly higher than
unity at the farthest end of our sample. This was one reason to choose the farthest distance limit 
for our study.
This redshift range gives the lower limit of group luminosity
$L_{gr} = 1.2 \times10^{10} h^{-2} L_{\sun}$ (red line in Fig.~\ref{fig:lgrdist}), in total 20855 groups.
To characterize galaxy groups, in our analysis we used data on galaxy groups (luminosity
$L_{gr}$, richness $N_{gal}$, and
velocity dispersions $\sigma_{\mathrm{v}}$) from the  \citet{2014A&A...566A...1T} catalogue.
We also  analysed   the properties of the BGGs.
In groups the brightest galaxies (BGGs) in $r$-band are defined as the brightest galaxies in a group.

Galaxies without any close neighbours are classified  as single galaxies.
Single galaxies may be the brightest galaxies of faint groups where
other group members are too faint to be included in the  SDSS spectroscopic sample.
Single galaxies may also be   systems in which one luminous galaxy is surrounded
by dwarf satellites \citep[called   hypergalaxies in][]{1974TarOT..48....3E}.
We included single galaxies in our study to better understand  the properties 
of the  BGGs of the faintest groups in our sample.
For our study it is important
that these galaxies do not have close neighbours of approximately the same luminosity.
In Sect.~\ref{sect:bgg} we show that the absolute magnitude limit
of BGGs is $M_r = -19.50$. Therefore, we used the absolute magnitude limited sample of 
single galaxies with the same magnitude limit. In total, our sample includes data on 43315 
single galaxies.

We note that SDSS spectroscopic data are known to be affected by  fibre collision effect,
which
means that approximately $6$~\% of the galaxies within a spectroscopic sample luminosity limit
remain without measured redshifts. \citet{2018MNRAS.481.2458D} performed a detailed analysis
on the influence of this effect to very  poor groups with two or three members.
They concluded that up to $10$~\% of galaxy pairs may be missing, but otherwise the
properties of groups remain unchanged. For details we refer to \citet{2018MNRAS.481.2458D}.
For our study this means that some single galaxies may actually be pair members.
Therefore, in order to see whether this   affects our results for the poorest groups,
we formed a test sample where, to mimic missing pairs, we added a subset ($10$~\%)   
of randomly chosen single galaxies to the sample of the BGGs of the poorest groups, 
and used these samples to estimate possible
biases in our analysis of the poorest groups, related to the fibre  collision effect.

\begin{figure*}
\centering
\resizebox{0.90\textwidth}{!}{\includegraphics[angle=0]{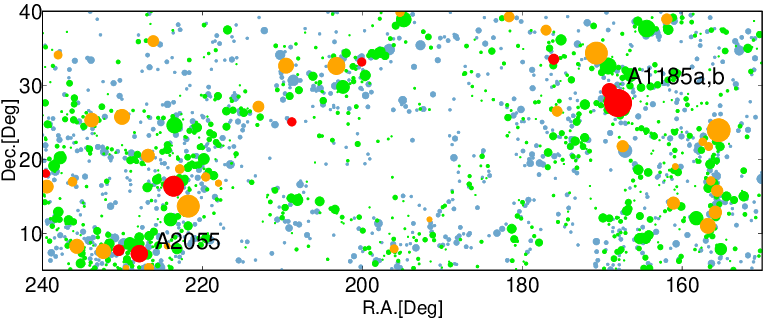}}
\caption{Sky distribution of groups of different richness.
The symbol sizes are proportional to the  richness of the groups.
Red denotes groups with  $L_{gr} \geq 43 \times10^{10} h^{-2} L_{\sun}$,
orange  groups with
$L_{gr} = 15 - 43 \times10^{10} h^{-2} L_{\sun}$; green   groups with 
 $L_{gr} = 2.5 - 15  \times10^{10} h^{-2} L_{\sun}$, 
and blue groups with $L_{gr} \leq 2.5 \times10^{10} h^{-2} L_{\sun}$.
Shown are the  Abell numbers for A2055 in the Hercules supercluster and for A1185 in the Leo supercluster.
}
\label{fig:radecid}
\end{figure*}

{\it Global environment of galaxies: Luminosity--density field}.
To characterize the  global environment of groups  we 
used
the luminosity--density field \citep{2012A&A...539A..80L}, calculated via a
smoothing kernel based on the $B_3$ spline function:
\begin{equation}
    B_3(x) = \frac{|x-2|^3 - 4|x-1|^3 + 6|x|^3 - 4|x+1|^3 + |x+2|^3}{12}.
\end{equation}
The details of the calculation of the density field using a $B_3$ spline kernel
can be found in \citet{2007A&A...476..697E} and \citet{2014A&A...566A...1T}.
We use a smoothing length of $8$~\Mpc\ to define the global luminosity--density
field,
and denote the  global luminosity--density as $D8$. 
Luminosity--density values are expressed in units of mean 
luminosity--density, $\ell_{\mathrm{mean}}$ = 
1.65$\cdot10^{-2}$ $\frac{10^{10} h^{-2} L_\odot}{(\vmh)^3}$. 
In the luminosity--density field, connected regions with the
highest luminosity--density above a threshold density of $D8 = 5.0$
are typically defined as  superclusters \citep[as in e.g.][]{2012A&A...539A..80L,
2014A&A...562A..87E, 2020A&A...641A.172E}. 
Regions of the
highest luminosity--density having $D8 \geq 7$ correspond to the high-density
cores of superclusters.
Superclusters only fill   $\approx 1$\% of the SDSS volume.
Most underdense regions between  superclusters, where $D8 < 1$, 
occupy approximately  $65$\% of the SDSS volume \citep{2019A&A...623A..97E}. 
To have a quick look at the sky distribution of groups 
of various luminosity, we show in Fig.~\ref{fig:radecid} the sky distribution of groups in our sample
in a sky area that  partly covers the Hercules and the Leo superclusters.

{\it Filaments in galaxy distribution}.
As one proxy of an environment of groups in the cosmic web, we used their location with respect to filaments. 
The data on galaxy filaments are   from the  filament catalogues 
by \citet{2014MNRAS.438.3465T} and \citet{2016A&C....16...17T}.
In these catalogues the  galaxy filaments were detected by 
applying a marked point process to the SDSS galaxy distribution (Bisous model).
For each galaxy, a distance from the nearest filament
axis was calculated. A galaxy
is considered to be  filament member  
if its distance from the nearest filament axis is within $0.5$~\Mpc,
as described in detail in \citet{2014MNRAS.438.3465T} and  \citet{2020A&A...641A.172E}.
Filaments cover a wide range of global densities, and represent an additional way to 
characterize the cosmic web.

{\it Galaxy data}.
In our study we obtained galaxy  data from the SDSS DR10 web 
page.\footnote{\url{http://skyserver.sdss3.org/dr10/en/help/browser/browser.aspx}}
We used  the following data to characterize BGGs of different star formation properties:
absolute $r$ and $g$ magnitude $M_r$ and $M_g$, stellar mass $M^\star$, stellar velocity dispersion $\sigma^\star$, 
$D_n(4000)$ index, and star formation rate $\log \mathrm{SFR}$.
The rest-frame galaxy colour index $(g - r)_0$   is defined as $(g - r)_0 = M_g - M_r$.

The stellar masses $M^\star$, star formation rates $\mathrm{(SFRs)}$, and stellar velocity dispersions $\sigma^{\mathrm{*}}$
are from
the MPA-JHU spectroscopic catalogue \citep{2004ApJ...613..898T, 2004MNRAS.351.1151B}.  
The parameters of galaxies were found using 
the stellar population synthesis models and fitting SDSS photometry and spectra 
with \citet{2003MNRAS.344.1000B} models. \citet{2003MNRAS.341...33K} gives the description how 
the stellar masses of galaxies were calculated. The   $\mathrm{SFR}$s 
were computed using the photometry and emission lines 
\citep[see][for details]{2004MNRAS.351.1151B} and \citet{2007ApJS..173..267S}. 
The stellar velocity dispersions of galaxies $\sigma^{\mathrm{*}}$ were found 
by fitting galaxy spectra and employing 
publicly available codes: the Penalized PiXel Fitting code
\citep[pPXF,][]{2004PASP..116..138C} 
and the Gas and Absorption Line Fitting code 
\citep[GANDALF,][]{2006MNRAS.366.1151S}.

Additionally, we used the data on the  $D_n(4000)$ index of galaxies from 
the MPA-JHU spectroscopic catalogue \citep{2004ApJ...613..898T, 2004MNRAS.351.1151B}.  
$\text{The }D_n(4000)$ index is the ratio of the average flux density
in the band $4000 - 4100 \angstrom$ to those in the band $3850 - 3950 \angstrom$. 
This index is correlated with the time passed since 
the most recent star formation event in a galaxy \citep{2003MNRAS.341...33K}.
We used the $D_n(4000)$ index of galaxies as calculated in \citet{1999ApJ...527...54B}.

\begin{figure}
\centering
\resizebox{0.46\textwidth}{!}{\includegraphics[angle=0]{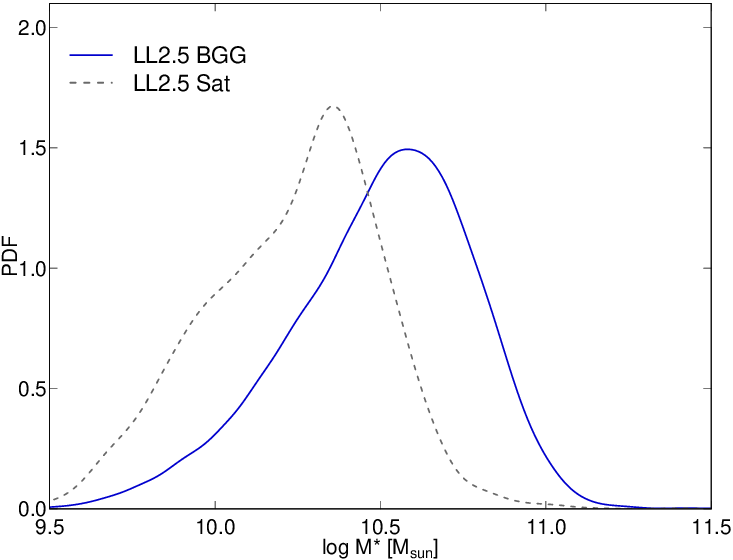}}
\caption{Distribution of stellar masses, $\log M^{\star}$,
for the BGGs and satellite galaxies in faint groups with 
$L_{gr} \leq 2.5 \times10^{10} h^{-2} L_{\sun}$.
The solid line correspond to BGGs and the dashed line to satellite galaxies in faint groups.
}
\label{fig:smd2.5s}
\end{figure}

The BGGs of groups also have   the highest stellar masses among group galaxies. For groups with luminosity 
$L_{gr} \geq 2.5 \times10^{10} h^{-2} L_{\sun}$ this was shown in \citet{2022A&A...668A..69E},
who compared the stellar masses and other properties of BGGs and satellite galaxies in groups.
As the   \citet{2022A&A...668A..69E} sample did not include groups with luminosities
$1.2 \times10^{10} h^{-2} L_{\sun} \leq L_{gr} < 2.5 \times10^{10} h^{-2} L_{\sun} $,
we show this for the faint groups in Fig.~\ref{fig:smd2.5s}, which presents 
the distribution of the stellar masses
of BGGs and satellite galaxies in faint groups with luminosities from this interval.

Using these data, we divided the BGGs 
into three classes according to their star formation properties 
\citep[see also][]{2020A&A...641A.172E}. 
The data of the galaxy populations are summarized in Table~\ref{tab:galpops}.
We use the following notation: Q for quenched galaxies with no active star formation 
($\log \mathrm{SFR} < -0.5$); 
RSF for red star-forming galaxies with $(g - r)_0 \geq 0.7$ and $\log \mathrm{SFR} \geq -0.5$; 
BSF for blue star-forming galaxies with $(g - r)_0 < 0.7$ and $\log \mathrm{SFR} \geq -0.5$.  

Here, a word of caution is needed. SDSS data are known to be affected by aperture bias, 
which may lead to underestimation of the SFR for highly star-forming galaxies.
We used the SFRs from \citet{2004MNRAS.351.1151B},   which take this  into
account. However, \citet{2017MNRAS.470..639G} showed that \citet{2004MNRAS.351.1151B} still underestimated this effect.
We did not analyse star formation properties of the BGGs in detail, and thus in the context of our study 
it is sufficient  to understand whether our results on the star formation properties 
of the BGGs are sensitive to the chosen limit to separate star-forming and quenched galaxies,
$\log \mathrm{SFR} = -0.5$. Therefore, we performed calculations also using the limits
$\log \mathrm{SFR} = -0.6$ and $\log \mathrm{SFR} = -0.4$.

\begin{table*}[ht]
\caption{BGG populations   used in this paper.}
\begin{tabular}{lll} 
\hline\hline  
(1)&(2)&(3)\\      
\hline 
Population & Abbr. &Definition  \\
\hline                                                    
Quenched galaxies with no star formation & Q &  $\log \mathrm{SFR} < -0.5$  \\
Red star-forming galaxies   & RSF &  $(g - r)_0 \geq 0.7$, $\log \mathrm{SFR} \geq -0.5$    \\
Blue star-forming galaxies   & BSF &  $(g - r)_0 < 0.7$, $\log \mathrm{SFR} \geq -0.5$    \\
\hline
\label{tab:galpops}  
\end{tabular}\\
\tablefoot{                                                                                 
Columns are as follows:
(1): BGG population;
(2): Abbreviation;
(3): Definition of a given population.
}
\end{table*}

\section{Classes of groups and dynamical properties of groups} 
\label{sect:met} 

To obtain the possible classes of groups, based on the properties of the groups, their BGGs,
and their environments, we applied unsupervised classification based
on multidimensional normal mixture modelling.
In this analysis we used the package {\it mclust} for classification and clustering
\citep{fraley2006} from the  {\it R}  statistical environment 
\citep{ig96}\footnote{\url{http://www.r-project.org}}. 
This package searches for an optimal model for the clustering of the data
among the models with varying shape, orientation, and volume, 
and determined the optimal number of  components in the data and the membership
of components (classification of the data). 
It studies a  finite mixture of distributions, 
in which each component is taken to correspond to a 
different class among groups. The
{\it mclust} package also calculates the uncertainty of the classification, which is defined as one minus the 
highest probability of a datapoint to belong to a component. 
It finds for each datapoint 
the probability to belong to a component. 
The mean uncertainty
for the full sample  is a statistical estimate of the reliability
of the results. The best solution for the components 
was chosen using the Bayesian information criterion (BIC). 

In our calculations we varied  input for {\it mclust}, and used 
different combinations of parameters, including group luminosity, luminosities, stellar masses and
star formation properties of BGGs, global luminosity--density at a group's location ($D8$), 
distance from filament axis ($D_{fil}$),  and other properties.
It is clear that the results of classification depend on input data, but a certain pattern 
emerged, in which groups were divided according to their luminosities, richness, star formation properties of BGGs,
and global luminosity--density  
in a persistent way. In Sect.~\ref{sect:results}  we present the results of classification step by step,
and show how the combination of parameters lead us to division of groups.

Then we study the properties of groups from different luminosity classes having different BGGs.
First we continue to analyse environments of groups, considering simultaneously global luminosity--density
and filament membership of groups from different luminosity classes and with different BGGs.
Then we compare the properties of groups of the same richness in different 
environments, and  
the properties and environment of the poorest groups with those of single galaxies.

Finally, we investigate dynamical properties of groups.
We analyse whether the  BGGs of groups lie at group centres or farther away.
In virialized clusters, galaxies follow the cluster potential well. Thus, we 
would expect that the main galaxies of rich groups and  clusters lie at the centres of 
groups (group haloes) and have small peculiar velocities
\citep{ost75,merritt84,malumuth92}. 
Therefore, the peculiar velocity  of the 
main galaxies in clusters is also an indication of the 
dynamical state of the cluster \citep[][and references therein]{coziol09}.  
However, the actual location of the BGGs depends on the orbit
of the BGG, the merging history of the clusters, and on other factors.
Several studies have shown, especially in the case of non-relaxed, multicomponent 
groups and clusters, that  the BGG often lie far from the centre. In the case of multicomponent groups the 
BGG may lie in the centre of one component, which may not be the main component \citep{2012A&A...540A.123E}.
To analyse the location of BGGs with regard to the group centres, we calculate the  normalized line-of-sight peculiar velocities 
of BGGs, 
 $V_{\mathrm{pec,n}} = |V_{\mathrm{pec}}|/{\sigma}_{v}$, 
where $|V_{\mathrm{pec}}|$ is the line-of-sight velocity difference between the BGG and group
centre, and ${\sigma}_{v}$ is the velocity dispersion of a group. We also calculate
the normalized offset of a BGG and group centre on the sky plane,
$D_{\mathrm{cen,n}} = D_{\mathrm{cen}}/r_{\mathrm{max}}$.
Here $D_{\mathrm{cen}}$ is the distance of the BGG from the group centre on the sky
plane, and $r_{\mathrm{max}}$ is the maximum size of a group in the sky plane.
We find these characteristics for high-luminosity groups only, as they are
poorly defined for poor groups \citep{2013MNRAS.434..784R}.
We also compare the stellar velocity dispersions of BGGs ($\sigma^{\mathrm{*}}$)
 and group velocity dispersions ($\sigma_{\mathrm{v}}$) in the case of high-luminosity groups.

\begin{figure}
\centering
\resizebox{0.46\textwidth}{!}{\includegraphics[angle=0]{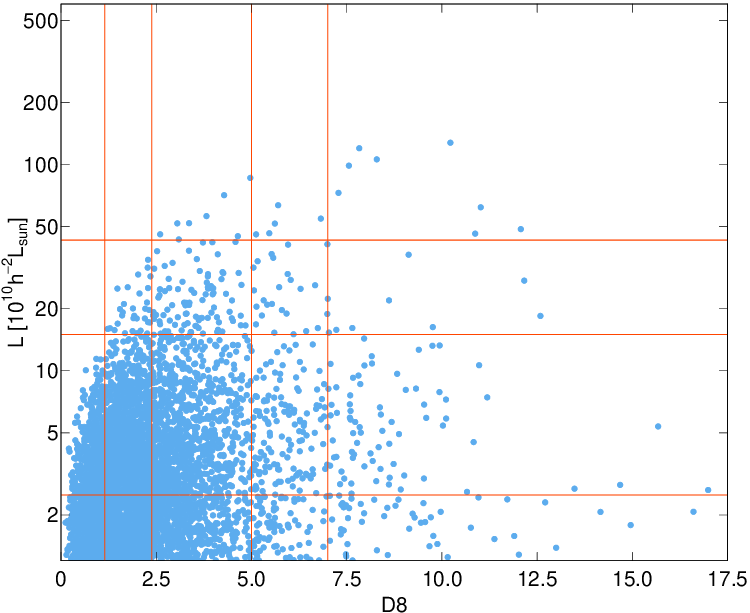}}
\caption{Luminosity of groups vs global luminosity--density $D8$.  The horizontal lines indicate the  group
luminosities $L^{thr}_{gr} = 2.5 \times10^{10} h^{-2} L_{\sun}$, 
$L^{thr}_{gr} = 15 \times10^{10} h^{-2} L_{\sun}$, and $L^{thr}_{gr} = 43 \times10^{10} h^{-2} L_{\sun}$.
The vertical lines show the  global luminosity--density limits $D8 = 1.15$, $D8 = 2.38$, $D8 = 5.0$, and $D8 = 7.0$ (see text). Only 
one-quarter of all  the groups, randomly chosen, are shown. 
}
\label{fig:lgrd8}
\end{figure}

\begin{figure}
\centering
\resizebox{0.46\textwidth}{!}{\includegraphics[angle=0]{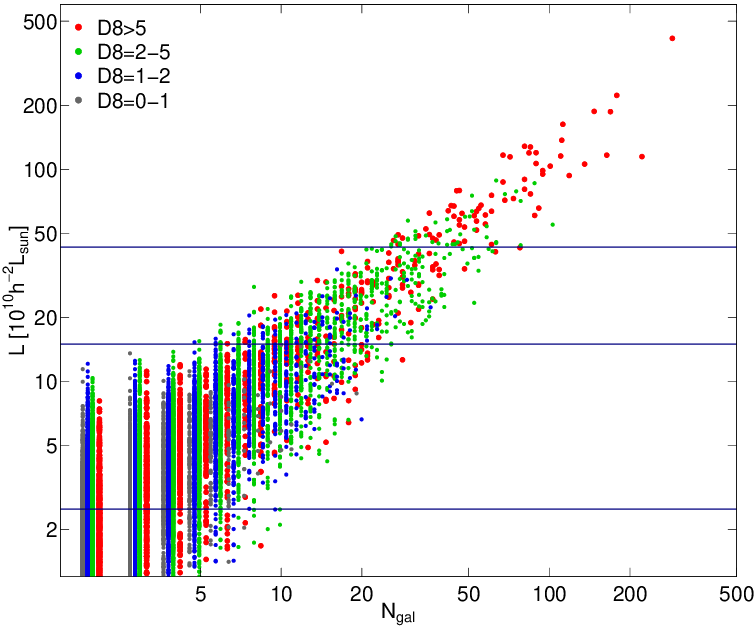}}
\caption{Luminosity of groups vs group richness. The colours denote groups in various global density D8 
regions, as shown in the figure. The horizontal lines are as in Fig.~\ref{fig:lgrd8}. 
}
\label{fig:lgrngal}
\end{figure}

\section{Results: Division of groups} 
\label{sect:results}

\begin{table}[ht]
\centering
\caption{Data on groups}
\begin{tabular}{lrrrr} 
\hline\hline  
(1)&(2)&(3)&(4)&(5)\\                                                          
\hline 
ID & $L_{gr}$ limit & $N_{\mathrm{gr}}$  & $N_{\mathrm{gal}}^{min}$  & $M^{med}_{gr}$\\
\hline                                                    
HL43     & > 43      &   107 & 25 & 176 \\
HL1543   & 15 - 43   &   619 &  5 &  40 \\
\hline                                                    
LL2.515  & 2.5- 15   & 11628 &  2 &  3\\
LL2.5    &  < 2.5    &  8501 &  2 &  0.8\\
\hline                                        
Single &           & 43315 & 1  &  \\
\hline                                        
\label{tab:grclass}  
\end{tabular}\\
\tablefoot{                                                                                 
Columns are as follows:
(1): Notation (ID);
(2): Luminosity limits of groups and clusters, $L_{gr}$, in units of $10^{10} h^{-2} L_{\sun}$;
(3): Number of groups/clusters in a given luminosity interval;
(4): Minimum richness of groups/clusters in a given luminosity interval, $N_{\mathrm{gal}}^{min}$;
(5): Median mass of groups/clusters in a given luminosity interval, $M^{med}_{gr}$,
in units of $10^{12} h^{-1} M_{\sun}$.
}
\end{table}

In this section we disentangle the results of group classification and 
analyse the properties of groups, their BGGs, and their environments step by step. 
Everywhere where we compare different populations we test the statistical significance
of the results using the Kolmogorov-Smirnov (KS) test. 
If  the estimated probability of rejecting the hypothesis
that the distributions are statistically similar (the $p$-value) $p \leq 0.01$
then the differences between distributions are {\it \textup{highly}} significant.
If not stated otherwise, 
the differences between the samples are found to be highly significant, and we do not present the 
results of the KS test in detail. 
We did not apply the  KS test when samples were too small, for example with fewer than 20 groups 
\citep[see e.g.  ][for the reliability of tests for samples of various size]{2013MNRAS.434..784R}.

\subsection{Global luminosity--density, group luminosity, and richness} 
\label{sect:lumr} 

We start with the analysis of division of groups according to the global luminosity--density, group luminosity, and richness.
Figures~\ref{fig:lgrd8} and 
\ref{fig:lgrngal} present the luminosity of groups versus global luminosity--density and versus their richness.
In Fig.~\ref{fig:lgrngal} the points are colour-coded according to the global luminosity--density region where the groups reside.
 We note that in this figure the richness values of very poor groups are shifted in order to avoid overlapping in the figure.
We also note   that even galaxy pairs and triplets have luminosity values in a rather wide range,
$1.2 \leq L_{gr} < 12 - 13 \times10^{10} h^{-2} L_{\sun}$, in all global environments.

The global luminosity--density field can be divided into regions of superclusters,
supercluster outskirts, and voids using characteristic threshold density as follows
\citep[see][and references therein]{2012A&A...539A..80L, 2021A&A...649A..51E, 2022A&A...668A..69E}.
We use  the threshold density of $D8 = 5$  to define the largest overdensity  regions - superclusters 
of galaxies. The threshold density of $D8 \approx 7$ delineates high-density cores of
rich superclusters \citep{2007A&A...464..815E}. 
Global luminosity--density  $D8 < 5$ outlines the outskirts of superclusters
and  the voids between superclusters.
The global lowest density
regions were called  the  watershed regions in \citet{2022A&A...668A..69E}.
The threshold density for the watershed regions is $D8 \leq 1$.

It is clear that the global luminosity--density, group luminosities, and group richness are correlated. 
First of all, as already found in \citet{2022A&A...668A..69E}, 
groups with luminosity 
$L_{gr} > 15\times10^{10} h^{-2} L_{\sun}$ are absent in the global lowest density
regions (watersheds).
The threshold density for watershed regions is $D8 = 1.15$ in the case of our present sample.
Therefore, groups can be divided into two main luminosity classes 
based on this threshold luminosity and density.
Figure~\ref{fig:lgrngal} shows that 
at this limit groups have richness values in a rather wide interval from 5 to 20 member galaxies.

In addition, {\it mclust} found subclasses of low- and high-luminosity groups,
with luminosity limits for high-luminosity groups $15 \leq L_{gr} < 43 \times10^{10} h^{-2} L_{\sun}$ and
$L_{gr} \geq 43 \times10^{10} h^{-2} L_{\sun}$.
Groups with $L_{gr} \geq 43 \times10^{10} h^{-2} L_{\sun}$ are only present at global
luminosity--density $D8 \geq 2.38$.
For low-luminosity groups the luminosity limits are $2.5 \leq L_{gr} < 15 \times10^{10} h^{-2} L_{\sun}$ and
 $L_{gr} < 2.5 \times10^{10} h^{-2} L_{\sun}$. These groups are located everywhere in the luminosity--density
field.
We present in Table~\ref{tab:grclass} a short summary of group properties in these luminosity
classes. 
In Table~\ref{tab:grclass} the subclasses are labelled in   order as HL43, HL1543 (HL15,
when taken together), LL2.515, and LL2.5 (LL15, when together).
We do not analyse group masses in this paper, as the masses of poor groups are not well
defined, but as many studies use masses instead of luminosity, we also provide in Table~\ref{tab:grclass}
 the median masses of groups in each class from
\citet{2014A&A...566A...1T}. The median mass of groups at luminosity
threshold ($L_{gr} = 15\times10^{10} h^{-2} L_{\sun}$) 
is $M^{med}_{gr} \approx 23 \times10^{12} h^{-1} M_{\sun}$.
We note that the masses of the groups differ by a factor of $10^{4}$ (Table~\ref{tab:grclass}).

Groups of the highest luminosity also are the richest, which is expected as
their luminosity is based on luminosities of galaxies in groups. These systems are also the most massive.
We also note   that groups with $L_{gr} \geq 100\times10^{10} h^{-2} L_{\sun}$ are present only
in superclusters or in their high-density cores (Fig.~\ref{fig:lgrd8}, 
19 clusters in our sample). 
Their luminosity and masses are comparable to those of rich clusters \citep[see e.g.][for
luminosity and masses of rich clusters in the Corona Borealis supercluster]{2021A&A...649A..51E}.
The sample of groups HL1543 
represent low-mass clusters and the richest groups. For clarity, in what follows we denote 
all high-luminosity groups (HL15) as clusters.
In doing so, we must keep in mind that the sample HL1543 
also includes   rich and luminous groups. The sample HL43 
represents clusters, including very rich clusters. 

Low-luminosity groups with $L_{gr} < 15 \times10^{10} h^{-2} L_{\sun}$ (LL15) are also less massive
and poor, with median mass $M^{med}_{gr} \approx 3 \times10^{12} h^{-1} M_{\sun}$ and less.
We note that the mass of the Local Group is 
$M_{MW+M31} \approx 4.2 \times10^{12} M_{\sun}$, which for $h = 0.7$ is similar to the median mass
of very poor groups 
\citep[see Table~\ref{tab:grclass} and ][for details and references]{2021PhRvD.103b3009L}.

The lowest luminosity groups are also the poorest, up to five member galaxies
(only $\sim$0.5~\% of them have 5--10 member galaxies); approximately $\sim$3~\% of the  LL2.515 groups 
have $10 - 20$ member galaxies. In the group sample by \citet{2022A&A...668A..69E}
the LL15 groups 
were all with fewer  than 
ten members. The difference comes from using other sample limits in the present study.
We  call these systems  poor and very poor groups.
For clarity, we use luminosity limits to denote the samples in the figures and tables.
In the next section we analyse the properties of the BGGs of groups from the different luminosity classes determined above.
This analysis is a second step in the division of groups into the various classes.

\begin{figure}
\centering
\resizebox{0.46\textwidth}{!}{\includegraphics[angle=0]{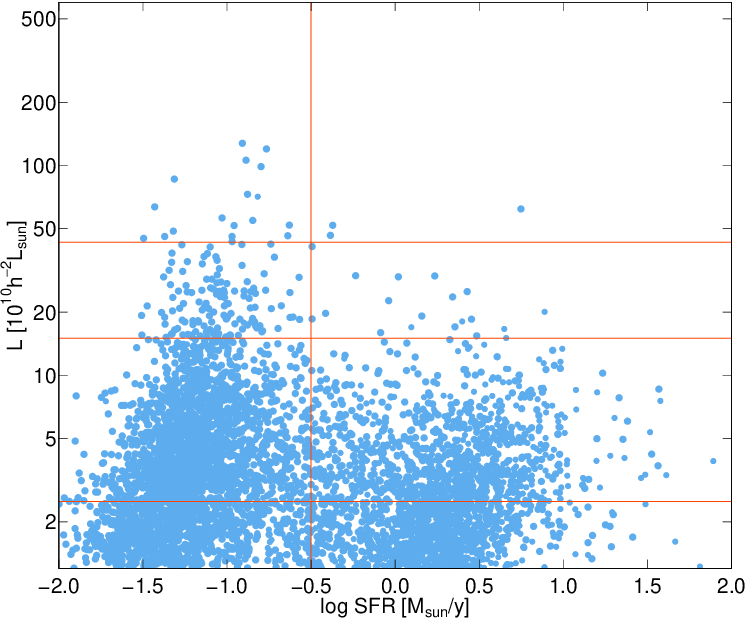}}
\caption{Luminosity of groups vs star formation rate, $\log \mathrm{SFR}$. 
The horizontal lines are as in Fig.~\ref{fig:lgrd8}. The vertical line shows the star formation rate limit
for star-forming and quenched galaxies, $\log \mathrm{SFR} = -0.5$. 
Only 
one-quarter of all  the groups, randomly chosen, are shown. 
}
\label{fig:lgrsfr}
\end{figure}

\begin{figure}
\centering
\resizebox{0.46\textwidth}{!}{\includegraphics[angle=0]{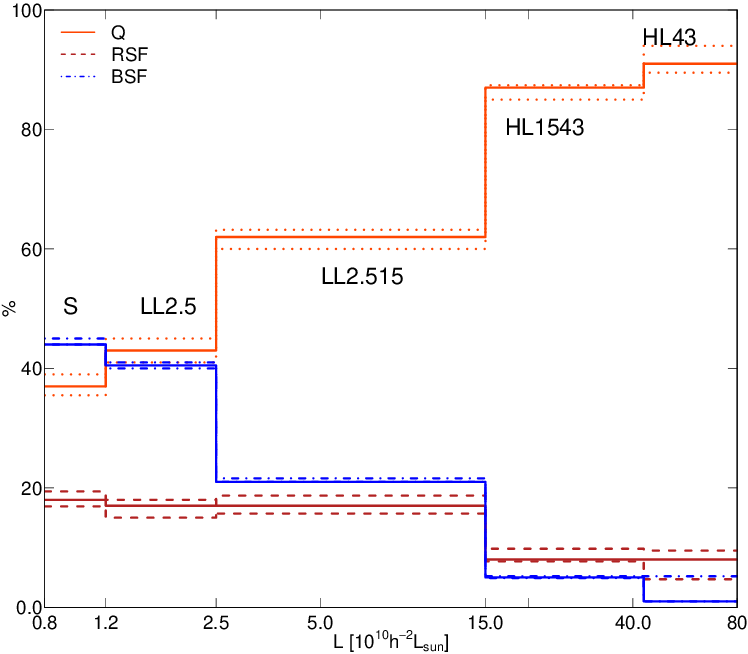}}
\caption{Percentages of groups with BGGs of different star formation properties vs group luminosity
for groups from different luminosity classes and for single galaxies (S), as shown in the figure. 
The red line shows the percentages of 
quenched BGGs and the  dark red dashed line shows the percentages of  RSF BGGs. 
The blue line shows the percentages of BSF BGGs.
The dotted, dashed, and dot-dashed lines show the changes in percentages if  SFR limits  $\log \mathrm{SFR} = -0.6$
and $\log \mathrm{SFR} = -0.4$ were used. The LL2.5 group sample in this figure 
is a test sample, to which is  added  $10$~\% of randomly chosen single galaxies.
}
\label{fig:bggfrac}
\end{figure}

\begin{figure*}
\centering
\resizebox{0.33\textwidth}{!}{\includegraphics[angle=0]{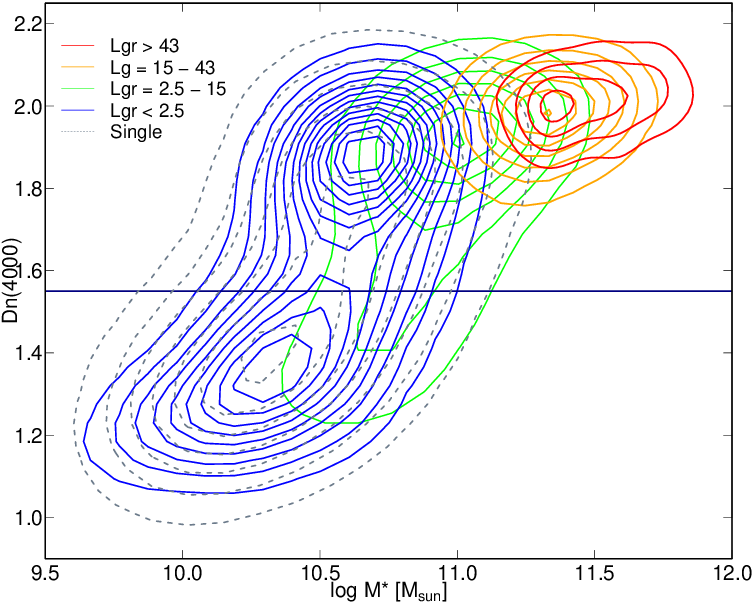}}
\resizebox{0.33\textwidth}{!}{\includegraphics[angle=0]{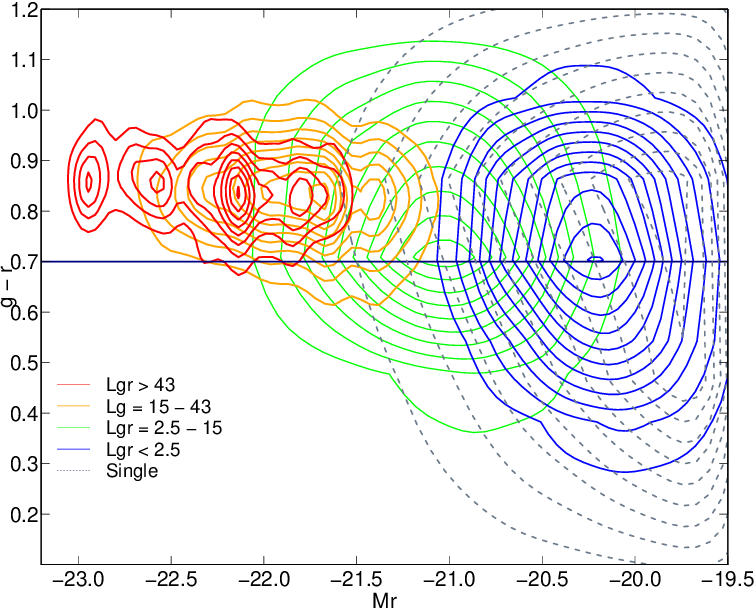}} 
\resizebox{0.32\textwidth}{!}{\includegraphics[angle=0]{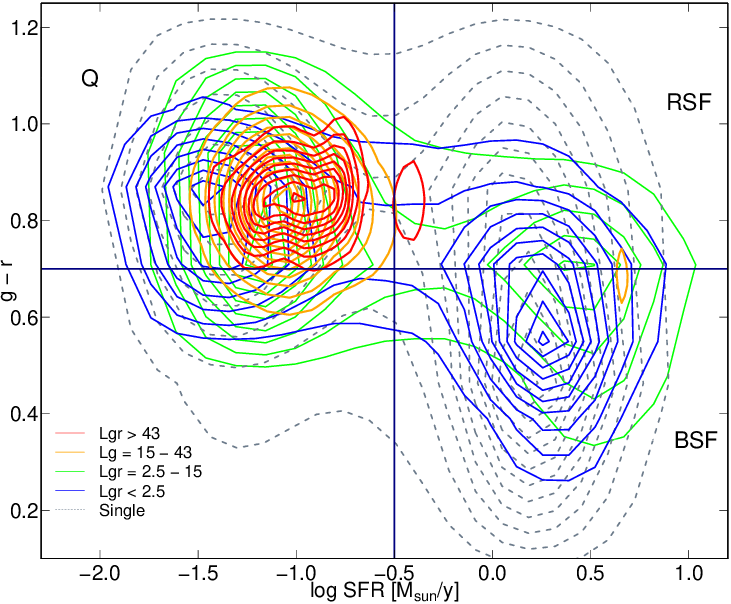}}
\caption{
Properties of BGGs and single galaxies. Left panel:  Stellar mass 
$\log M^{\star}$  vs $D_n(4000)$ index.
Middle panel:  $(g - r)^0$ colour index vs absolute magnitude in $r$ colour
(colour--magnitude relation). Right panel: Star formation rate
$\log \mathrm{SFR}$ vs $(g - r)^0$ colour index.
The line colours correspond to groups of different luminosity and to single galaxies, as 
shown in the panels. The horizontal line in the left panel separates quenched and star-forming galaxies
at $D_n(4000) = 1.55$. In the middle and right panels the horizontal line shows the division between red and blue galaxies
at $(g - r)^0 = 0.7$. The vertical line in the  right panel shows the separation of quenched and star-forming galaxies
at $\log \mathrm{SFR} = -0.5$.
}
\label{fig:smdngrsingle}
\end{figure*}

\begin{table}[ht]
\centering
\caption{Percentages of different BGGs in group classes}
\begin{tabular}{lrrr} 
\hline\hline  
(1)&(2)&(3)&(4)\\      
\hline 
ID &  $F_{Q}$  & $F_{RSF}$ & $F_{BSF}$  \\
\hline                           
HL43     & 0.91{\raisebox{0.5ex}{\tiny$^{+0.03}_{-0.02}$}}&0.08{\raisebox{0.5ex}{\tiny$^{+0.03}_{-0.02}$}} & 0.01 \\
HL1543   & 0.87{\raisebox{0.5ex}{\tiny$^{+0.01}_{-0.01}$}}&0.08{\raisebox{0.5ex}{\tiny$^{+0.02}_{-0.01}$}} & 0.05{\raisebox{0.5ex}{\tiny$^{+0.00}_{-0.02}$}}\\
\hline  
LL2.515  & 0.62{\raisebox{0.5ex}{\tiny$^{+0.01}_{-0.02}$}}&0.17{\raisebox{0.5ex}{\tiny$^{+0.02}_{-0.01}$}} & 0.21{\raisebox{0.5ex}{\tiny$^{+0.00}_{-0.01}$}} \\
LL2.5    & 0.45{\raisebox{0.5ex}{\tiny$^{+0.02}_{-0.02}$}}&0.16{\raisebox{0.5ex}{\tiny$^{+0.01}_{-0.01}$}} & 0.39{\raisebox{0.5ex}{\tiny$^{+0.01}_{-0.00}$}} \\
LL2.5*   & 0.42{\raisebox{0.5ex}{\tiny$^{+0.01}_{-0.01}$}}&0.17{\raisebox{0.5ex}{\tiny$^{+0.02}_{-0.01}$}} & 0.41{\raisebox{0.5ex}{\tiny$^{+0.01}_{-0.00}$}} \\
\hline                                                                                                    
Single &  0.37{\raisebox{0.5ex}{\tiny$^{+0.01}_{-0.02}$}}& 0.18{\raisebox{0.5ex}{\tiny$^{+0.01}_{-0.01}$}} & 0.44{\raisebox{0.5ex}{\tiny$^{+0.01}_{-0.00}$}}  \\
\hline                                        
\label{tab:bggclass}  
\end{tabular}\\
\tablefoot{                                                                                 
Columns are as follows:
(1): Notation (ID, see text). LL2.5* denotes the average of the sample to which we added $10$~\% of randomly selected 
single galaxies to the LL2.5 sample;
(2): Percentage of groups with a quenched BGG having $\log \mathrm{SFR} \leq -0.5$ (Q);
(3): Percentage of groups with a red star-forming BGG (RSF).
(4): Percentage of groups with a blue star-forming BGG (BSF).
In Cols. 2--4 we show the  changes in percentages if we had used SFR limits  $\log \mathrm{SFR} = -0.6$
and $\log \mathrm{SFR} = -0.4$.
}
\end{table}

\subsection{Star formation properties of BGGs} 
\label{sect:bgg} 

As a first look at the BGG properties for groups of different luminosity, 
we show in Fig.~\ref{fig:lgrsfr} the luminosities of groups versus the star formation properties of their brightest
galaxies, $\log \mathrm{SFR}$. 
Table~\ref{tab:bggclass} shows the percentage of groups with quenched and star-forming BGGs among
groups of different luminosity classes.
The percentages of BGGs of different star formation properties in groups of different class, and for single
galaxies are also shown in Fig.~\ref{fig:bggfrac}.
In Table~\ref{tab:bggclass} and in Fig.~\ref{fig:bggfrac} we also show
the changes in percentages of BGGs with different star formation properties if we used SFR limits  $\log \mathrm{SFR} = -0.6$
and $\log \mathrm{SFR} \leq -0.4$.
In Fig.~\ref{fig:smdngrsingle}  we present 
the stellar mass $\log M^{\star}$ -- $D_n(4000)$ index plane,
colour-magnitude diagram, and star formation rate--colour diagram for BGGs of groups from 
different luminosity classes.

In these figures and table  we see a clear difference in the properties of BGGs of groups and clusters,
and a difference  between each subclass.
Clusters have mostly red BGGs with no active star formation (approximately $90$~\% of clusters).
Among the  19 richest clusters only two have star-forming BGGs.
In Fig.~\ref{fig:smdngrsingle} the BGGs of the HL43 groups (red contours) 
form a compact cloud
in parameter space with high stellar masses, high values of the $D_n(4000)$ index,
red colours, and low star formation rates. 

In contrast,  Figs.~\ref{fig:lgrsfr} and ~\ref{fig:smdngrsingle} show that the star formation properties of BGGs 
of poor groups and single galaxies  are  bimodal. These trends 
are seen best in Table~\ref{tab:bggclass},  Table~\ref{tab:grlummed}, and   Fig.~\ref{fig:bggfrac}. 
At the lowest luminosity there is a nearly 
equal amount of quenched (Q) galaxies and  BSF galaxies,
both about $40 - 45$~\%. These proportions change systematically with group luminosity, 
reaching $91$~\% and $1$~\%, respectively, for HL43 groups. 
Table~\ref{tab:bggclass} and Fig.~\ref{fig:bggfrac} show that if we change the classification
threshold  of the star formation rate,
then the changes in percentages of groups with BGGs of various type are very small, typically $2$~\% or less.
The corresponding percentages for single galaxies and BGGs of LL2.5 groups have close values. 
Therefore, it is possible
that some single galaxies are misclassified because of fibre collisions, and they belong to LL2.5 population.

The proportion of intermediate, RSF galaxies varies less, 
being about 1/6 for poor groups and 1/12 for clusters. 
RSF galaxies are considered to be  galaxies in transition from the blue to the red cloud. Below
we  study groups with RSF BGGs separately.
Single galaxies follow the same trends as BGGs of LL2.5 groups, but with 
a larger scatter. 
There are also blue galaxies with low star formation rates. They form $\approx 2$\%
of all BGGs in our sample, and we do not analyse groups with such BGGs separately.

In summary, there is a clear difference of BGG properties between the BGGs of 
clusters and groups. Therefore, the properties of BGGs, together with luminosity and the location
in the cosmic web, support the division of groups and clusters into two main classes, HL15 and LL15.
The  BGGs of each subclass (HL1543 and HL43, and LL2.515 and LL2.5) differ in the 
percentage of star-forming BGGs; lower luminosity groups have a higher percentage of blue star-forming
groups than higher luminosity groups and clusters.
We also note that the threshold luminosity, $L_{gr} \approx 15 \times10^{10} h^{-2} L_{\sun}$,
is approximate, and our results hold around this luminosity value.

\begin{table}[ht]
\centering
\caption{Median values of luminosity of groups and clusters with different BGGs.}
\begin{tabular}{lrrr} 
\hline\hline  
(1)&(2)&(3)&(4)\\      
\hline 
ID &  \multicolumn{3}{c}{$L_{gr}$}  \\
          & $Q$  & $RSF$  & $BSF$   \\
\hline                                                    
HL43       & 60.1&52.4 & 49.0 \\
HL1543     & 20.4&19.1 & 18.4 \\
\hline   
LL2.515    & 4.5 &4.1   &3.6 \\
LL2.5      & 1.8 &1.8  & 1.7 \\
\hline                                        
\label{tab:grlummed}  
\end{tabular}\\
\tablefoot{                                                                                 
Columns  are as follows:
(1):  Notation (ID);
(2--4): Median values of group and cluster luminosity, $L_{gr}$
(in $10^{10} h^{-2} L_{\sun}$),
 for groups  from  a given luminosity range with  quenched (Q), 
red star-forming (RSF), and blue star-forming (BSF) BGGs.
}
\end{table}

\section{Correlations with the global luminosity--density and filament membership for
 groups and clusters with different BGGs} 
\label{sect:bggtrans}

\begin{figure*}
\centering
\resizebox{0.46\textwidth}{!}{\includegraphics[angle=0]{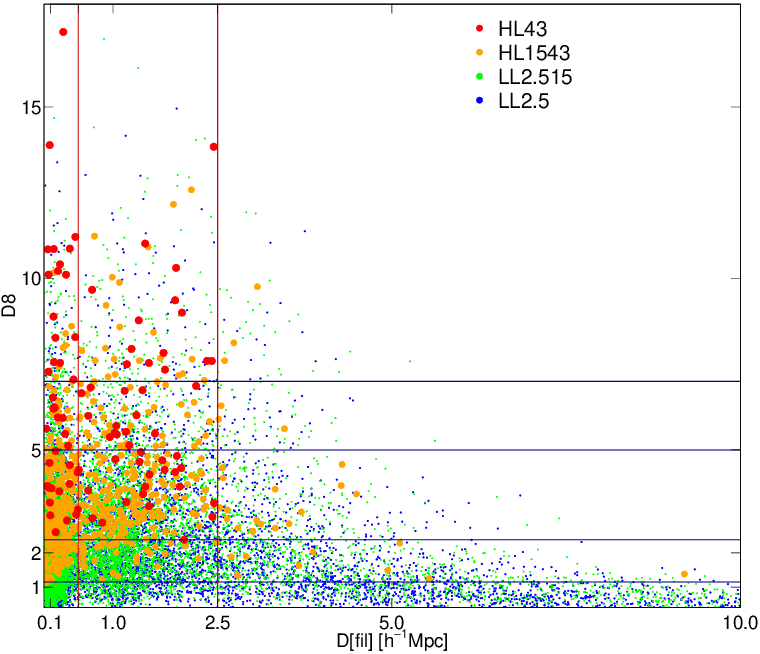}}
\resizebox{0.477\textwidth}{!}{\includegraphics[angle=0]{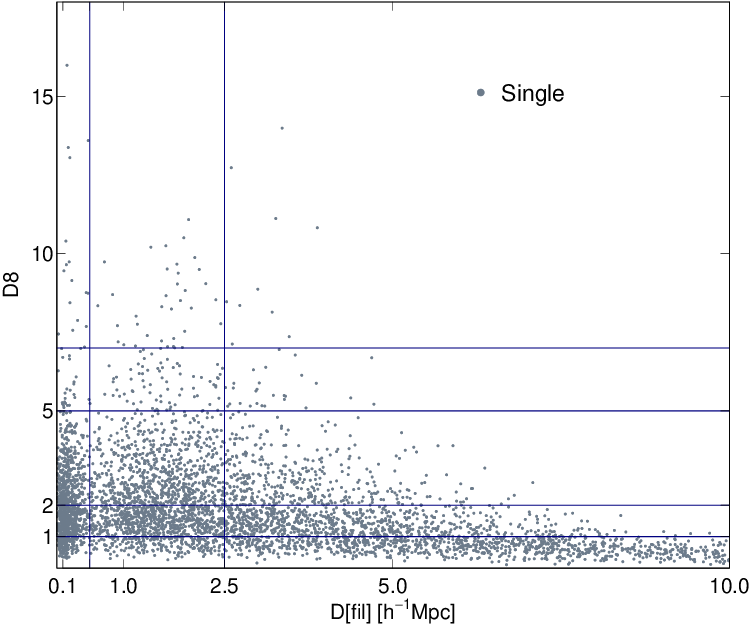}}
\caption{Global density $D8$ vs distance from the nearest filament axis $D_{fil}$ for groups (left panel) and for
single galaxies (right panel). The colours denote groups in 
four luminosity classes. The horizontal lines indicate global luminosity--density limits 
$D8 = 1.15$, $D8 = 2.38$, $D8 = 5.0$, and $D8 = 7.0$. The vertical 
lines show the characteristic distances from the nearest filament axis,
$D_{fil} = 0.5$~\Mpc\ and $D_{fil} = 2.5$~\Mpc.
}
\label{fig:d8dfil}
\end{figure*}

\begin{figure}
\centering
\resizebox{0.46\textwidth}{!}{\includegraphics[angle=0]{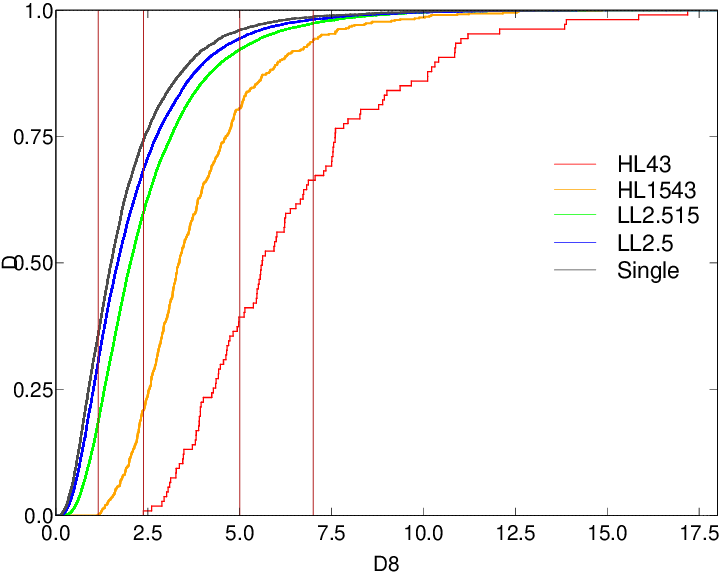}}
\caption{
Cumulative distribution of global luminosity--density at the location of groups. The line colours
correspond to the luminosity classes, as shown in the figure. 
The vertical lines indicate the  global luminosity--density limits $D8 = 1.15$, $D8 = 2.38$, $D8 = 5.0$, and $D8 = 7.0$.
}
\label{fig:d8all}
\end{figure}


\begin{figure}
\centering
\resizebox{0.464\textwidth}{!}{\includegraphics[angle=0]{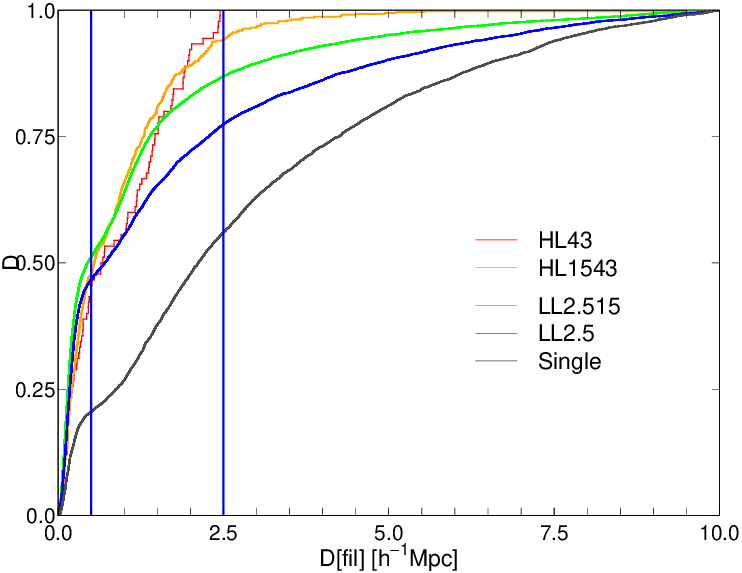}}
\caption{
Cumulative distributions of distances of groups from the nearest filament axis 
for groups of different luminosity. The line colours
correspond to the luminosity classes, as shown in the figure. 
The vertical lines show the characteristic distances from the nearest filament axis,
$D_{fil} = 0.5$~\Mpc\ and $D_{fil} = 2.5$~\Mpc.
}
\label{fig:dfilbgg}
\end{figure}

\begin{table}[ht]
\centering
\caption{Median values of the global luminosity--density at group location, $D8$.}
\begin{tabular}{lrrr} 
\hline\hline  
(1)&(2)&(3)&(4)\\      
\hline 
ID &   \multicolumn{3}{c}{$D8$}  \\
   &                 $Q$  & $RSF$  & $BSF$   \\
\hline                                                    
HL43       & 5.62&5.93 & 5.45 \\
HL1543     & 3.37&3.08 & 2.91 \\
\hline  
LL2.515    & 2.08&1.98  &1.91 \\
LL2.5      & 1.72&1.75 & 1.61 \\
\hline                                                    
Single &            1.55&1.52 & 1.40 \\
\hline                                        
\label{tab:d8med}  
\end{tabular}\\
\tablefoot{                                                                                 
Columns are as follows:
(1):  Notation (ID);
(2--4): Median values of the global luminosity--density at group location, $D8$,  for
groups from a given luminosity range with   quenched (Q), 
red star-forming (RSF), and blue star-forming (BSF) BGGs.
}
\end{table}


\begin{table}[ht]
\centering
\caption{Groups and their BGGs in various global luminosity--density $D8$ regions.}
\begin{tabular}{lrrrrr} 
\hline\hline  
(1)&(2)&(3)&(4)&(5)&(6)\\      
\hline 
ID & $D8$ limits & $F_{gr}$  & $F_{Q}$ & $F_{RSF}$ & $F_{BSF}$   \\
\hline                                                    
HL43     &    107     &  &    &  &  \\
\hline                                                    
       & 2.38 - 5 &   0.40 & 0.95 & 0.05&  \\
       &  5 - 7   &   0.26 & 0.83 & 0.14&0.03  \\
       & > 7      &   0.34 & 0.90 & 0.10&   \\
\hline                                        
\hline                                                    
HL1543     &  619   & &    &  &  \\
\hline                                                    
       & 1.15 - 5 &   0.81 & 0.87 & 0.09& 0.04 \\
       &  5 - 7   &   0.13 & 0.81 & 0.11&0.08  \\
       & > 7      &   0.06 & 0.89 & 0.03& 0.08  \\
\hline                                                    
LL2.515     & 11628   &  &    &  &  \\
\hline                                                    
       & 0.0 - 1.15 &   0.18 & 0.57 & 0.17& 0.26 \\
       & 1.15 - 2.38 &  0.41 & 0.62 & 0.17& 0.21 \\
       & 2.38 - 5   &   0.32 & 0.64 & 0.16& 0.20 \\
       & > 5        &   0.08 & 0.64 & 0.17& 0.18  \\
\hline                                                    
LL2.5    &   8501    &   &    &  &  \\
\hline                                                    
       & 0.0 - 1.15 &   0.30 & 0.44 & 0.15& 0.41 \\
       & 1.15 - 2.38 &  0.38 & 0.45 & 0.16& 0.39 \\
       & 2.38 - 5   &   0.26 & 0.47 & 0.16& 0.37 \\
       & > 5        &   0.06 & 0.49 & 0.16& 0.35  \\
\hline                                                    
Single &    43315        &  &    &  &  \\
\hline                                                    
       & 0.0 - 1.15 &   0.37 & 0.35 & 0.17& 0.48 \\
       & 1.15 - 2.38 &  0.37 & 0.39 & 0.18& 0.44 \\
       & 2.38 - 5   &   0.21 & 0.41 & 0.19& 0.41 \\
       & > 5        &   0.04 & 0.44 & 0.19& 0.36  \\
\hline                                        
\label{tab:grdenlimfrac}  
\end{tabular}\\
\tablefoot{                                                                                 
Columns are as follows:
(1):  Notation (ID);
(2): Global luminosity--density limits, $D8$, and number of groups/clusters and single galaxies
in a given luminosity interval;
(3): Percentage of groups with a given luminosity in a given density interval;
(4): Percentage of groups in a given density interval with quenched BGG having $\log \mathrm{SFR} \leq -0.5$;
(5): Percentage of groups in a given density interval with a red star-forming BGG (RSF).
(6): Percentage of groups in a given density interval with a blue star-forming BGG (BSF).
}
\end{table}

\begin{table}[ht]
\centering
\caption{Median values of the distances from the nearest filament axis, $D_{fil}$,
for filaments with length $L \geq 3$~\Mpc.}
\begin{tabular}{lrrr} 
\hline\hline  
(1)&(2)&(3)&(4)\\      
\hline 
ID &   \multicolumn{3}{c}{$D_{fil}$}  \\
   &                 $Q$  & $RSF$  & $BSF$   \\
\hline                                                    
HL43      & 0.65&1.0 & 1.0 \\
HL1543    & 0.56&0.41 & 0.56 \\
\hline  
LL2.515   & 0.44&0.42  &0.48 \\
LL2.5     & 0.61&0.74 & 0.80 \\
\hline                                                    
Single &           1.98&2.18 & 2.32 \\
\hline                                        
\label{tab:dfilmed}  
\end{tabular}\\
\tablefoot{                                                                                 
Columns are as follows:
(1):  Notation (ID);
(2--4): Median values of the distances from the nearest filament axis, 
$D_{fil}$ (in \Mpc), for groups and clusters from  a given luminosity range with  quenched (Q), 
red star-forming (RSF), and blue star-forming (BSF) BGGs.
}
\end{table}


\begin{table}[ht]
\centering
\caption{Groups and their BGGs in filaments, in filament outskirts, and far from filaments}
\begin{tabular}{lrrrrr} 
\hline\hline  
(1)&(2)&(3)&(4)&(5)&(6)\\      
\hline 
ID & $D_{fil}$ limits & $F_{gr}$  & $F_{Q}$ & $F_{RSF}$ & $F_{BSF}$   \\
\hline                                                    
HL43     &   107     &   &    &  &  \\
\hline                                                    
       & < 0.5     &   0.45 & 0.90 & 0.10&  \\
       & 0.5 - 2.5 &   0.54 & 0.88 & 0.10&0.02  \\
\hline                                        
\hline                                                    
HL1543     &  619  & &    &  &  \\
\hline                                                    
       & < 0.5     &   0.48 & 0.85 & 0.10&0.05\\
       & 0.5 - 2.5 &   0.46 & 0.88 & 0.07&0.05  \\
       & > 2.5     &   0.06 & 0.82 & 0.12& 0.06  \\
\hline                                                    
LL2.515     &  11628   &  &    &  &  \\
\hline                                                    
       & < 0.5     &   0.50 & 0.62 & 0.17&0.21\\
       & 0.5 - 2.5 &   0.35 & 0.65 & 0.16&0.20  \\
       & > 2.5     &   0.14 & 0.54 & 0.23& 0.27  \\
\hline                                                    
LL2.5    &   8501    &   &    &  &  \\
\hline                                                    
       & < 0.5     &   0.45 & 0.47 & 0.15&0.38\\
       & 0.5 - 2.5 &   0.30 & 0.46 & 0.16&0.38  \\
       & > 2.5     &   0.25 & 0.42 & 0.20& 0.44  \\
       & > 2.5v    &   0.14 & 0.41 & 0.23& 0.46  \\
\hline                                        
Single &   40494        &  &    &  &  \\
\hline                                                    
       & < 0.5     &   0.20 & 0.41 & 0.17&0.42   \\
       & 0.5 - 2.5 &   0.33 & 0.40 & 0.18&0.42   \\
       & > 2.5     &   0.47 & 0.35 & 0.23& 0.48  \\
       & > 2.5v    &   0.27 & 0.34 & 0.26& 0.51  \\
\hline                                        
\label{tab:grdfillimfrac}  
\end{tabular}\\
\tablefoot{                                                                                 
Columns are as follows:
(1):  Notation (ID);
(2): Distance from a nearest filament axis, $D_{fil}$~\Mpc\
(2.5v marks groups with  $D_{fil} \geq 2.5$~\Mpc\ and $D8 \leq 1.15$, i.e. extreme void environment, 
far from filaments), and number of groups (clusters or single galaxies) in a given distance interval;
(3): Percentage of groups of a given luminosity at a given $D_{fil}$ interval;
(4): Percentage of groups at a given $D_{fil}$ interval with quenched BGG having $\log \mathrm{SFR} \leq -0.5$;
(5): Percentage of groups at a given $D_{fil}$ interval with red star-forming BGG (RSF).
(6): Percentage of groups at a given $D_{fil}$ interval with blue star-forming BGG (BSF).
}
\end{table}

To characterize the environment of groups in the cosmic web, we now  use 
the distance of a group or cluster 
from the nearest filament axis ($D_{fil}$)  together with the global luminosity--density field. 
We only use data on filaments with lengths
greater than $3$~\Mpc, because they are  more reliable \citep{2020A&A...639A..71K, 2021A&A...649A..51E}.
We plot in Fig.~\ref{fig:d8dfil} the location of groups of different luminosity on the
$D8$--$D_{fil}$ plane.
In this figure we indicate the  threshold density for superclusters and their high-density cores,
$D8 = 5$ and $D8 = 7$, as well as the threshold densities as density limits for HL15 and HL1543
clusters, $D8 = 1.15$ and $D8 = 2.38$.  
As mentioned above, groups and single galaxies can be considered as members of filaments if their 
distance from
the nearest filament axis is $D_{fil} \leq 0.5$~\Mpc. We show this value in Fig.~\ref{fig:d8dfil}.
We also show the value $D_{fil} = 2.5$~\Mpc; this is the maximum value of $D_{fil}$
for the sample of rich clusters. In what follows we   denote regions with $0.5 < D_{fil} \leq 2.5$~\Mpc\
as filament outskirts.

We show in Fig.~\ref{fig:d8all}  the cumulative distribution of the global luminosity--density $D8$ at group location.
For comparison with the lowest luminosity groups, we also show this distribution   for single galaxies.
Figure~\ref{fig:dfilbgg} shows the distributions of distances from the nearest
filament axis for groups and clusters of different luminosity classes.
In our further analysis we analyse the properties of groups and their BGGs
in filaments, in filament outskirts, and far from filaments, which is defined
as regions with $D_{fil} > 2.5$~\Mpc.


Figures~\ref{fig:d8dfil} and \ref{fig:dfilbgg} show  that
 approximately $45 - 50$~\% of all groups and clusters are filament members
with $D_{fil} \leq 0.5$~\Mpc.
Moreover, almost all HL1543 clusters ($96$~\%)
have $D_{fil} \leq 2.5$~\Mpc. Therefore, they lie in filaments or in filament outskirts. 
Clusters, especially rich clusters, are extended objects surrounded by regions of influence
and may have radii
up to several  megaparsecs \citep{2014A&A...566A...1T, 
2020A&A...641A.172E, 2021A&A...649A..51E}.
To be connected to a filament, it is enough for a cluster that this
filament reaches the outer parts of a cluster or its region of influence.
Therefore, all clusters in HL43 can be considered to be connected to a filament.

In the lowest global
luminosity--density regions with $1.15 < D8 < 2.38$ (the density limits 
at which there are no very rich clusters)
even 78~\% of poor clusters lie at
distances from the nearest filament axis $D_{fil} \leq 1$~\Mpc,
which is approximately the size of poor groups \citep{2014A&A...566A...1T}.
This is in agreement with understanding that groups form at the intersections 
of filaments. At low global luminosity--density regions
even in filaments clusters cannot become very rich; 
the very rich clusters
can only be seen in  filaments at $D8 > 2.38$.
This may be related to higher connectivity of clusters in superclusters, as was shown
in \citet{2020A&A...641A.172E} for clusters in the A2142 supercluster and in
low-density region around it. To verify this assumption
for a large sample of groups and clusters a separate study is needed.

Now we compare environments of groups
and clusters with BGGs of different  star formation properties. 
Table~\ref{tab:d8med}  presents median values of $D8$  at the location of groups
and clusters with different BGGs, and Tables~\ref{tab:grdenlimfrac},
\ref{tab:dfilmed}, and \ref{tab:grdfillimfrac} show
how groups with different BGGs
from each luminosity range are spread between regions of different global luminosity--density
and at various distances from the filament axis.

From these tables we see that
high- and low-luminosity groups do not populate the luminosity--density field in the same way.
The richest clusters (HL43)
have minimal threshold density at their location, $D8 = 2.38$. A total of  
$40$\% of rich clusters lie at supercluster outskirts, at global luminosity--density $D8 = 2.38 - 5$,
and $60$\% lie in superclusters ($34$\% in supercluster high-density cores with $D8 \geq 7$). The 
L1543 groups  
populate mostly supercluster outskirts or even voids ($80$\% of such groups), but they are absent in extreme 
void environments ($D8 < 1.15$). Only $6$\% of these groups lie in supercluster high-density cores.
In contrast, poor (LL15) groups
lie mostly in low global luminosity--density environments. Less than  $10$\%
of these lie in superclusters, and this percentage decreases for the poorest groups and
single galaxies. 

At all group luminosities, groups with red quiescent  BGGs lie in higher density environments than groups
with blue star-forming  BGGs. In addition,
Table~\ref{tab:dfilmed} tells  that, on average, groups and clusters 
with quenched  BGGs have lower values of the distances
from the  filament axis than groups with RSF  BGGs, and, in  turn, groups with RSF BGGs
have lower values of the distances from the filament axis than groups with BSF  BGGs.

At the same time, 
Table~\ref{tab:grdenlimfrac} shows that the percentages of groups from a given luminosity range
having  quiescent, RSF or BSF BGGs do not change with global luminosity--density.
The same can be seen from Table~\ref{tab:grdfillimfrac}: the percentages of groups with different BGGs
do not change with the group's distance from the filament axis. Even among the poorest  lowest luminosity
groups in extreme void regions these percentages are the same as among groups in filaments
and in filament outskirts. 
This suggests that group and cluster properties are modulated by their location in the cosmic
web, but the properties of the BGGs in them are mostly determined by processes
within group or cluster dark matter halo.

In contrast to groups, $20$~\% of single galaxies are 
members of filaments with $D_{fil} \leq 0.5$~\Mpc,  and altogether $47$~\% of single galaxies
lie far from the nearest filament axis with  $D_{fil} > 2.5$~\Mpc\
(Tables~\ref{tab:dfilmed} and \ref{tab:grdfillimfrac}).
In addition, far from filaments the percentage of both red and blue star-forming BGGs of the lowest luminosity groups and 
single galaxies increases and the percentage of quenched BGGs and single galaxies decreases.
In agreement with this, the median distance from the nearest filament axis 
increases. These trends are weak or absent in the case of clusters.

\begin{figure*}
\centering
\resizebox{0.46\textwidth}{!}{\includegraphics[angle=0]{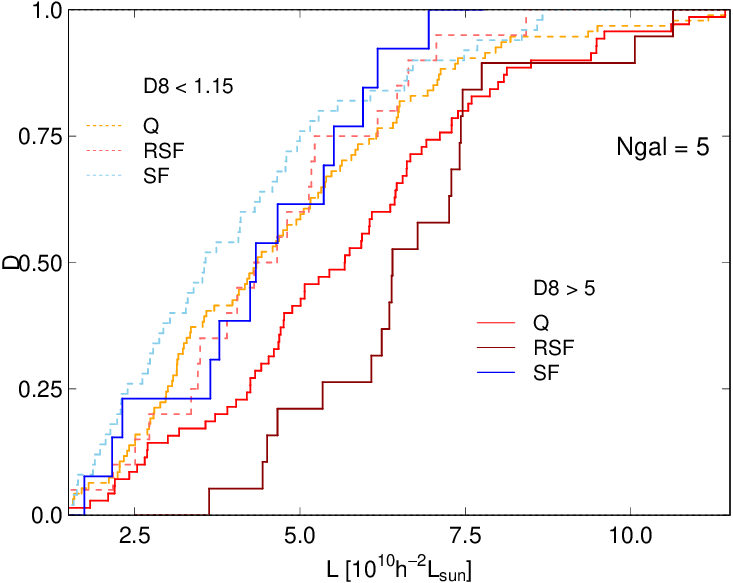}}
\resizebox{0.47\textwidth}{!}{\includegraphics[angle=0]{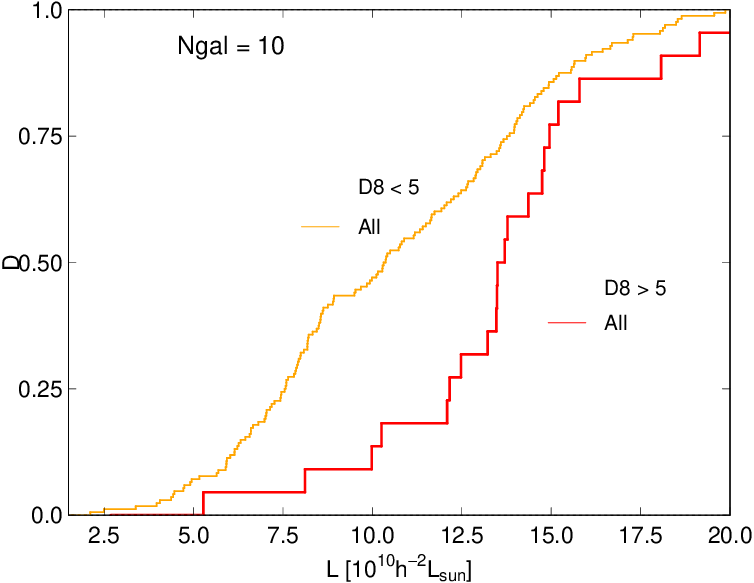}}
\caption{Cumulative distributions of group luminosities in high and low global luminosity--density
regions for groups with $N_{gal} = 5$ with different BGGs (left panel) and for groups with
$N_{gal} = 10$ (right panel).
The line colours correspond to groups with different BGGs in high- and low-density 
environments, as shown in the figure.
}
\label{fig:ng5hilo}
\end{figure*}

\begin{figure}
\centering
\resizebox{0.46\textwidth}{!}{\includegraphics[angle=0]{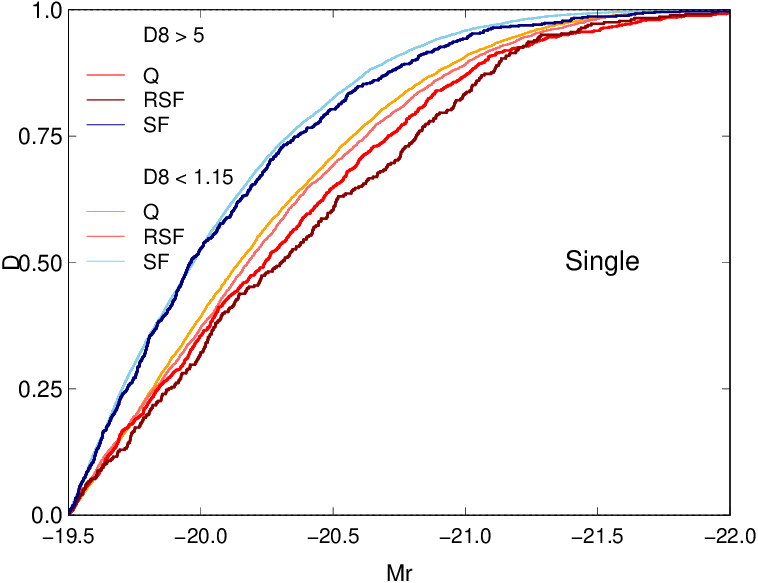}}
\caption{Cumulative distributions of luminosities of single galaxies $M_r$ in high and low global luminosity--density
regions. 
Luminosity--density limits are shows in each panel.
}
\label{fig:singlermagd8dcum}
\end{figure}

{\it Luminosities of poor groups of the same richness in various environments.}
Above we showed that
the richest and most luminous clusters reside in superclusters. Clusters with quenched BGGs have higher 
global luminosity--density values at their location than clusters with star-forming BGGs. 
Next we  compare luminosities of poor groups of the same richness 
in high-density (supercluster) and low-density (void) environments.
In this test we varied the richness of the groups; in Fig.~\ref{fig:ng5hilo} 
we show the results for $N_{gal} = 5$ and $N_{gal} = 10$.
For groups with $N_{gal} = 5$ we show distributions of luminosities 
for groups with different BGGs. In the case of groups with $N_{gal} = 10$, 
owing to the small number of groups, we do not divide them according to the BGG properties.
We also compare the luminosities of single galaxies with different star formation properties
in superclusters and in voids (Fig.~\ref{fig:singlermagd8dcum}).

Our calculations show that groups of the same richness in superclusters have higher luminosities
than in  voids. 
Groups with $N_{gal} = 5$  with red (quenched and star-forming)
BGGs are more luminous in superclusters than in voids (Fig.~\ref{fig:ng5hilo}). 
In voids, groups with quenched and RSF BGGs have statistically similar luminosities,
while in superclusters groups with $N_{gal} = 5$ with RSF BGGs are more luminous than groups with quenched BGGs.
Luminosities of poor groups with BSF BGGs and $N_{gal} = 5$ are statistically the same in all
 environments.

Figure~\ref{fig:singlermagd8dcum} shows that a similar result also holds  for single galaxies:
single galaxies with the same star formation properties have higher luminosities in higher
global luminosity--density environments. This is statistically highly significant for 
all galaxy types. 
We note that the spread of luminosities among groups with $N_{gal} = 5$ with BSF BGGs,
both in superclusters and in voids is smaller than the spread of luminosities among groups
with red BGGs, and also smaller than among single galaxies. 
This is also seen in Fig.~\ref{fig:singlermagd8dcum}. 

Interestingly, the differences in group luminosities in superclusters
and voids are the largest at a richness value of approximately $N_{gal} = 10$,
although the overall spread of group luminosities is larger at lower richness values (Fig.~\ref{fig:lgrngal}).
Groups with $N_{gal} = 10$ in superclusters have median luminosity
$L_{gr} \approx 13.6\times10^{10} h^{-2} L_{\sun}$, while  in voids the median luminosity of such
groups is $L_{gr} \approx 10.3\times10^{10} h^{-2} L_{\sun}$. 
This may be related to the low percentage of groups with BSF BGGs among group with  $N_{gal} = 10$. 
These groups showed no difference in luminosities in superclusters and in voids. This finding 
gives additional support to the division of the group sample at approximately this richness.

Finally, we compared luminosities of groups with $N_{gal} = 5$
and $N_{gal} = 10$ in and near filaments,
and in filaments,  in filament outskirts, and far from filaments. Our analysis shows trends that
groups of the same richness in filaments  are more luminous than groups far from filaments, but these
results were statistically not significant. Therefore, we do not present the corresponding figures.

For poor groups these trends have also been found   in earlier studies.
\citet{2003A&A...401..851E} and \citet{2005A&A...436...17E} showed, using data from
observations and simulations, that poor groups near rich clusters are more luminous
and have higher values of velocity dispersions than groups far from rich clusters.
\citet{2012A&A...545A.104L} and \citet{2017A&A...597A..86P} demonstrated  that
groups of the same richness are more luminous in superclusters than in voids.
Moreover, \citet{2017A&A...597A..86P} found that BGGs of poor groups in filaments 
have a higher probability to be of early type than the  BGGs of groups outside filaments.
Our findings in this study confirm these results.


\section{Dynamical properties of groups of different luminosity and BGGs} 
\label{sect:grprop} 

\begin{figure*}
\centering
\resizebox{0.46\textwidth}{!}{\includegraphics[angle=0]{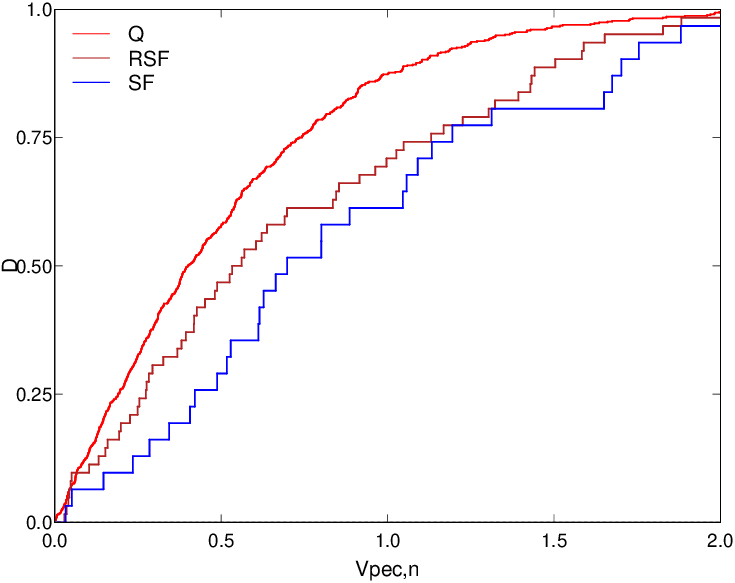}}
\resizebox{0.464\textwidth}{!}{\includegraphics[angle=0]{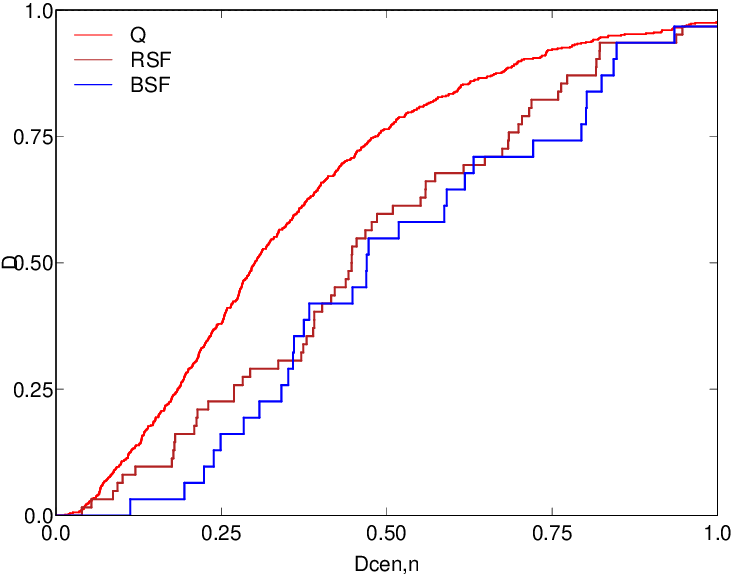}} 
\caption{
BGG location with respect to the cluster centre (sample HL15). Left panel: Normalized line-of-sight velocity,
$V_{\mathrm{pec,n}}$. Right panel: Normalized distance from the cluster centre on the sky plane,
$D_{\mathrm{cen,n}}$.
The line colours correspond to clusters with different BGGs, as 
shown in the panels. The red line denotes clusters with quenched (Q) BGGs, the dark red line denotes clusters with
RSF BGGs, and the  blue line denotes   clusters with BSF  BGGs.
}
\label{fig:bggcen}
\end{figure*}

\begin{figure}
\centering
\resizebox{0.46\textwidth}{!}{\includegraphics[angle=0]{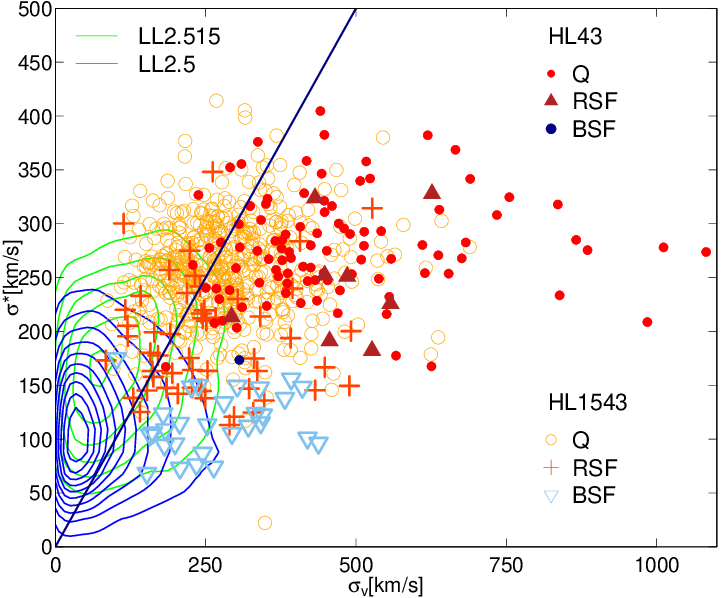}}
\caption{BGG stellar velocity dispersion $\sigma^{\mathrm{*}}$ vs group velocity dispersion $\sigma_{\mathrm{v}}$.
The symbol colours correspond to groups of different luminosity, as shown in the  panels. 
The line correspond to $\sigma^{\mathrm{*}} = \sigma_{\mathrm{v}}$.
}
\label{fig:sssvall}
\end{figure}

In this section we compare the dynamical properties of groups and clusters with different BGGs.
For this purpose, we first analyse the location of the BGGs of clusters
with respect to the cluster centre (Fig.~\ref{fig:bggcen}).
Then we compare the stellar velocity dispersions of BGGs $\sigma^{\mathrm{*}}$
 and cluster velocity dispersion $\sigma_{\mathrm{v}}$ (Fig.~\ref{fig:sssvall}).
We do not analyse poor groups in detail as in these groups the properties are not reliably defined.
As above, we divide the  BGGs according to their star formation properties.

We
present for clusters  the normalized line-of-sight  velocities and projected to the sky distances 
from the cluster centre in Fig.~\ref{fig:bggcen}.
As the number of rich clusters with star-forming BGGs is small, and distributions are very similar
for HLG43 and HLG1543 clusters, we do not show the  distributions for  the two subclasses of
clusters separately. 
Figure~\ref{fig:bggcen} shows that  quenched BGGs 
are located closer to the cluster centre than star-forming BGGs.
In the sky plane, the median values of normalized distances $D_{\mathrm{cen,n}}$ for quenched
BGGs  are  $D_{\mathrm{cen,n}} \approx 0.30$,
while for clusters with RSF BGGs $D_{\mathrm{cen,n}} \approx 0.45$.
For clusters with  blue
star-forming BGGs
the values of $D_{\mathrm{sky,n}}$ are the largest, $D_{\mathrm{cen,n}} \approx 0.47$.  

Along the line of sight the median value of normalized distances $V_{\mathrm{pec,n}}$
for clusters with quenched BGGs 
$V_{\mathrm{pec,n}} \approx 0.44$, and for clusters with star-forming BGGs are 
$V_{\mathrm{pec,n}} \approx 0.55$ and $V_{\mathrm{pec,n}} \approx 0.70$
for red and blue star-forming BGGs, correspondingly.
\citet{2012A&A...540A.123E} found that high values of the distances of BGGs may indicate
that clusters are multimodal, and the brightest galaxy may  be located near the centre
of a component in a cluster, but not in the central component. The detailed analysis of substructure
of rich groups and clusters with different BGGs could be a topic for another study. 
In this paper we conclude that clusters with star-forming BGGs 
are dynamically different from clusters with quenched BGGs, as their BGGs have not yet reached the
cluster centre.

Next we analyse the velocity dispersions $\sigma_{\mathrm{v}}$ of groups from different luminosity
classes, having different BGGs (Fig.~\ref{fig:sssvall}). As mentioned, velocity dispersions of
poor groups are not reliable, they are plotted for comparison.
Figure~\ref{fig:sssvall} shows that $\sigma^{\mathrm{*}}$ values are the
highest for quenched BGGs, RSF BGGs have intermediate stellar velocity dispersions, and
 BSF BGGs have the lowest stellar velocity dispersions.
The BGG stellar velocity dispersion $\sigma^{\mathrm{*}}$ is proportional to the group and cluster
velocity dispersion  $\sigma_{\mathrm{v}}$ in a wide range of velocity dispersions, and flattens
at high end values of  $\sigma_{\mathrm{v}}$.
Among clusters with the highest values of  $\sigma_{\mathrm{v}}$ 
($\sigma_{\mathrm{v}} \gtrsim 600$ and $\sigma^{\mathrm{*}} \gtrsim 200$) 
there are very rich, multicomponent 
clusters in superclusters. Their high values of  $\sigma_{\mathrm{v}}$ and multimodality may be related; 
this is an interesting topic for a separate study.

At the lower end of group velocity dispersions simulations predict very few groups
\citep[see Fig. 4 in ][]{2021MNRAS.507.5780M}. In Fig.~\ref{fig:sssvall}
this area is populated by clusters with star-forming BGGs.
We may assume that   these BGGs will also  eventually be quenched, and for this reason in simulations
the number of such BGGs is very small.
Clearly, more detailed analysis of the relations between the properties and formation of clusters and 
their BGGs is needed in order to understand better the formation of clusters and their BGGs
\citep[see also recent studies by][on the dynamical evolution of clusters 
and their BGGs in the  Illustris simulation]{2022ApJ...938....3S}.

\begin{figure}
\centering
\resizebox{0.47\textwidth}{!}{\includegraphics[angle=0]{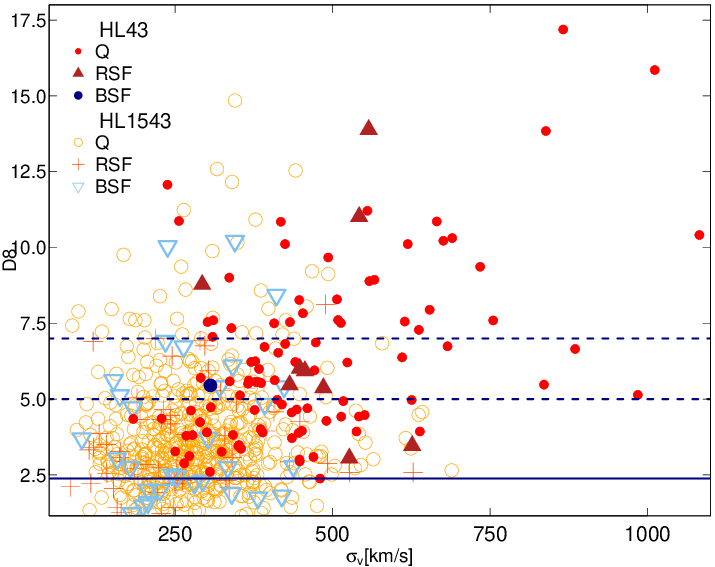}}
\caption{
Cluster velocity dispersion $\sigma_{\mathrm{v}}$ vs global luminosity--density 
at the location of clusters, separately for
clusters with BGGs of different star formation properties,
as shown in the plot. Horizontal lines mark global luminosity--density limits 
$D8 = 2.38$, $D8 = 5.0$, and $D8 = 7.0$.
}
\label{fig:d8sigmav}
\end{figure}

\begin{figure}
\centering
\resizebox{0.47\textwidth}{!}{\includegraphics[angle=0]{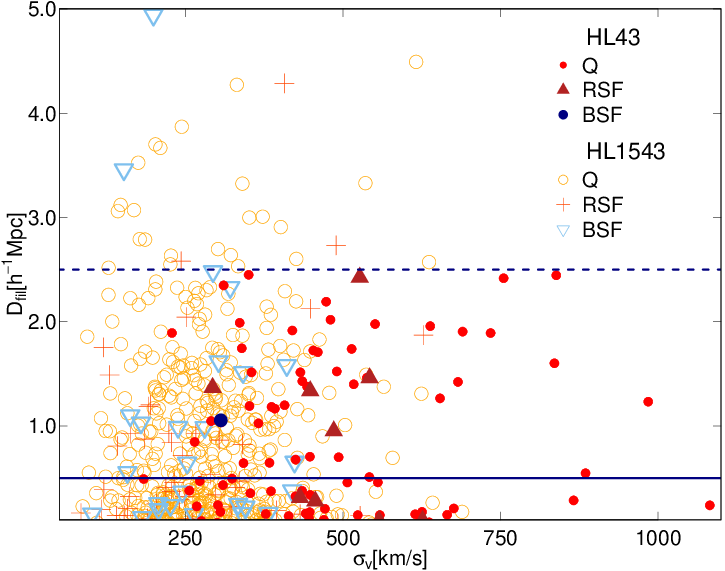}}
\caption{
Cluster velocity dispersion 
$\sigma_{\mathrm{v}}$ vs the distance from filament axis, separately for
clusters with BGGs of different star formation properties,
as shown in the plot. The horizontal 
lines show characteristic distances from the nearest filament axis,
$D_{fil} = 0.5$~\Mpc\ and $D_{fil} = 2.5$~\Mpc.
}
\label{fig:dfilsigmav}
\end{figure}

Figure~\ref{fig:d8sigmav} shows for high-luminosity clusters that
clusters with the highest velocity dispersions $\sigma_{\mathrm{v}}$ mostly populate 
superclusters and their high-density cores with $D8 \geq 7$. Out of 27 groups with
$\sigma_{\mathrm{v}} \geq 600$ (flattened part in
group velocity dispersion $\sigma_{\mathrm{v}}$ vs the  BGG stellar velocity dispersion
figure, Fig.~\ref{fig:sssvall}) only 8 groups  lie outside of superclusters,
while 13 of these clusters lie in high-density cores of superclusters. This again suggests that the growth 
of clusters and the evolution of their BGGs in superclusters and elsewhere are different,
and that  the properties of BGGs are more related to processes in the inner core of clusters and not to
the large-scale environment.
In Fig.~\ref{fig:dfilsigmav} we see that there is no large difference in how
clusters with the highest velocity dispersions $\sigma_{\mathrm{v}}$ populate 
filaments. Nine of these clusters lie at the filament axis, while 12 are located in filament 
outskirts.

\section{Summary and discussion}
\label{sect:discussion} 

We analysed the properties of galaxy groups and clusters 
with BGGs of different star formation properties in the cosmic web, characterized
by the luminosity--density field and filament membership. 
Our study showed that groups and clusters
can be divided into two main classes according to their luminosity, BGG properties, minimal
global luminosity--density at their location, and filament membership. We call these classes   groups and
clusters.
The threshold luminosity of groups and clusters
between these main classes is approximately $L_{gr} = 15 \times10^{10} h^{-2} L_{\sun}$
(mass $M_{gr} \approx 23 \times10^{12} h^{-1} M_{\sun}$, and richness $N_{\mathrm{gal}} \approx 10$)
(Figs.~\ref{fig:lgrd8} - \ref{fig:bggfrac}).

The BGGs of groups and clusters have different star formation properties.
In approximately $90$~\%  of clusters the brightest galaxies are  red, with no active star formation.
In contrast, 
 in poor groups and among single galaxies only $\approx 40 - 60$~\% of BGGs
are red and quenched, and
$\approx 60$~\% of the BGGs of poor groups and
single galaxies  are either red or blue star-forming galaxies.
Clusters lie in filaments or the filament outskirts in superclusters or the supercluster
outskirts, and are absent in void regions between superclusters where $D8 \leq 1$.
The richest clusters with luminosity  $L_{gr} \geq 100 \times10^{10} h^{-2} L_{\sun}$
can only be found in superclusters with $D8 \geq 5$.
In contrast, poor low-luminosity groups and single galaxies 
lie everywhere in the global luminosity--density field, including in void environments,
in filaments, and also far from filaments.
Clusters with different BGGs have different dynamical properties.
Groups of the same richness are more luminous in superclusters than in voids. This difference is
the largest approximately at the richness limit between groups and clusters, $N_{\mathrm{gal}} \approx 10$.
Next we briefly discuss  our results in the context of group evolution in the cosmic web.
We start with a short comparison with previous work.

\subsection{Comparison with previous work. }
\label{sect:comp}

Our study covered the properties and the location in the cosmic web of a wide range
of galaxies and galaxy systems, from single galaxies to rich clusters. We start
the comparison from single galaxies. Single galaxies may be isolated galaxies, that is,  
galaxies that do not have close bright galaxies according to certain criteria
\citep{2006A&A...449..937S, 2014A&A...566A...1T, 2016A&A...588A..79L}.

We found that approximately one-third of single galaxies are already quenched, with no active
star formation. This is the same percentage as found in \citet{2020A&A...641A.172E}
in underdense regions around the supercluster SCl~A2142.
This percentage is higher than found among isolated galaxies in, for example,
the AMIGA sample of isolated galaxies, where up to $14$~\%   are 
elliptical and lenticular galaxies \citep{2006A&A...449..937S}. However, we analyse a sample
in a wide range of environments, and our sample does not include very faint galaxies,
and this explains the difference. Among single galaxies the percentage of star-forming
galaxies is higher in the lowest global density environments, where galaxy population 
is mostly formed by faint star-forming galaxies \citep{2016MNRAS.458..394B}.

Small groups of galaxies (galaxy pairs and triplets), similar to our LL2.5 groups,
were analysed in \citet{2018MNRAS.481.2458D} and in \citet{2020MNRAS.493.1818D}. 
The authors showed that galaxies in very small groups tend to have higher star formation rates,
and they lie in lower density environments than galaxies in groups with four or more
member galaxies, in qualitative agreement of our study.
 
Galaxy groups have been studied in many papers, but a different choice of groups 
complicates the detailed comparison of the various results. Next we briefly describe some studies of
groups. Recently, a statistically complete
optically selected sample of 53 nearby groups within $80$~Mpc 
(the Complete Local-Volume Groups Sample, CLoGS), was analysed in a series of papers
\citep{2017MNRAS.472.1482O, 2018A&A...618A.126O, 2018MNRAS.481.1550K,
2022MNRAS.510.4191K, 2022MNRAS.515.1104L, 2022MNRAS.516.5487L}.
The authors estimate that the X-ray bright groups have masses in the range 
$M_{500} \approx 0.5-5 \times10^{13} M_{\sun}$. Therefore,  the CLoGS sample
 includes groups  in a mass range approximately similar to the high-mass end of our 
LL2.515 groups and the low-mass end of HL1543 groups.
In 18 groups from this sample the brightest galaxies are of early-type 
\citep{2022MNRAS.515.1104L}. This is lower percentage than the percentage of
quenched galaxies in our LL2.515 and HL1543 groups, but the difference can be
explained with a different selection criteria of the groups. The poorest groups in this sample
contain four bright galaxies \citep{2017MNRAS.472.1482O}.
Therefore, some CLoGS groups may be comparable to low-luminosity LL2.515 groups 
in our study. Among these groups the percentage of star formation  is higher,
and this may explain the lower percentage of early-type galaxies among  CLoGS groups.

Recent papers by \citet{2016MNRAS.458.2762G} and \citet{2018MNRAS.475.2787G}
are dedicated to the study of the brightest galaxies in X-ray groups over a wide range 
of masses, $M_{200} \approx 10^{12.8} - 10^{14} M_{\sun}$, and redshifts,
$0.04 \leq z \leq 1.3$.  These groups correspond approximately to our
LL2.515 and HL15 groups, but   from a much wider redshift range.
\citet{2016MNRAS.458.2762G} and \citet{2018MNRAS.475.2787G} found that the percentage of
star-forming BGGs in groups increases   towards lower group masses and higher redhshifts.
Our findings are in agreement with the first trend in star formation properties
from these studies. 
\citet{2020A&A...635A..36G} showed that in low-mass groups BGGs have higher 
peculiar velocities. In our study we found that in HL15 groups
star-forming BGGs have higher peculiar velocities than quenched BGGs. 
Star-forming BGGs have lower stellar masses than quenched BGGs, and in this respect our results
are in qualitative agreement with those by \citet{2020A&A...635A..36G}.

In very rich clusters the percentage of star-forming brightest galaxies
(in clusters these are called as the brightest cluster galaxies, BCGs) 
is very low, as found also in studies
of the brightest galaxies in groups and clusters by, for example, \citet{2019MNRAS.487.3759C} and \citet{2022MNRAS.512.2758O}. 
They found that in low-mass groups the percentage of star-forming BGGs may be up to
$30$~\%, while in rich clusters less than $10$~\% of BGGs are star forming.
\citet{2019MNRAS.487.3759C} mention that some star-forming BGGs may have red colours.
They suggest that the colours of these galaxies may be affected by the presence of dust
in them.

\citet{2017A&A...597A..86P} and \citet{2020A&A...639A..71K} 
compared the properties of the BGGs of groups from the SDSS sample
near the filament axis and outside of filaments. For very poor groups
comparable to our LL2.515 groups, \citet{2020A&A...639A..71K} found that
the properties of the BGGs in and near filaments are statistically similar.
They concluded that the properties of BGGs in very poor groups are mainly
determined by the local group environment.
\citet{2017A&A...597A..86P} used galaxy groups based on the volume-limited 
group sample drawn from the SDSS data to show that groups in superclusters
have higher masses, and their BGGs have higher stellar masses than groups
in low-density regions. They also compared the properties of the BGGs in filaments
and far from filaments in detail, and showed that BGGs  in groups in 
filaments are more luminous and their central galaxies have
higher stellar mass, redder colours, and lower star formation rates than those outside of filaments.
The masses of groups based on SDSS data in superclusters and elsewhere 
have been recently compared by \citet{2023arXiv230906251S}. 
\citet{2023arXiv230906251S} found that groups in superclusters have, on average, higher masses
than groups in low-density regions, in agreement with our results.

\citet{2011ApJ...736...51E} found that in the richest superclusters 
of the Sloan Great Wall, groups with early-type BGGs have more uniform
distribution than groups with late-type BGGs.
For a small sample of groups and clusters in the Corona Borealis supercluster
\citet{2021A&A...649A..51E} determined a concordance between the properties
of clusters and groups in their spheres of influence: groups within
the spheres of influence of clusters with a higher percentage
of quenched galaxies also have higher percentage of quenched galaxies
than groups farther away.

\subsection{Growth of groups and clusters in the cosmic web. }
\label{sect:grevol} 

According to the current cosmological paradigm, the evolution
of the structure  started in the early
inflationary stage of the evolution of the Universe.
Superclusters that embed rich clusters are formed
by medium- and large-scale perturbations, which combine in similar
overdensity phases, and amplify the growth of small-scale
perturbations \citep{2011A&A...534A.128E}.
Voids are regions in space where medium- and large-scale density
waves combine in similar underdensity phases, and suppress the
growth of galaxy-scale perturbations.  
The luminosity limits of groups and clusters in the global luminosity--density field 
come from the way   density waves of different wavelength combine to form the cosmic web.
Numerical simulations suggest that seeds of present-day rich
clusters of galaxies with first-generation stars were already present  at
redshift $z \geq 30$ \citep{2005MNRAS.363..379G, 2005MNRAS.363..393R}. These perturbations
were the basis of the skeleton of the cosmic web.

The seeds of galaxies are small-scale density perturbations that  grew via 
merging and the accretion of smaller structures \citep{2011A&A...534A.128E}. 
Medium- and large-scale perturbations
modulate the evolution of small-scale perturbations.
Galaxies can form everywhere in the cosmic web where large-scale density
perturbations in combination with small-scale overdensities are high
enough \citep{2021arXiv210602672P, Repp:2019wl}.

Rich clusters lie in or near superclusters  in filaments or in filament outskirts, and 
grow by infall of single galaxies and poor groups  along filaments
\citep{2009MNRAS.400..937M, 2014Natur.513...71T, 2019A&A...623A..97E, 2021A&A...652A..94E, 2022arXiv221205984M,
2023MNRAS.525.4685S}.  Single galaxies and poor groups 
are richer and more luminous in superclusters; this phenomenon has been called 
the  environmental enhancement of poor groups by \citet{2003A&A...401..851E}
and \citet{2005A&A...436...17E}. 
A similar property of groups has been found in \citet{2012A&A...545A.104L} and in \citet{2017A&A...597A..86P}. 
This is an additional factor that  enhances the growth of rich clusters
in superclusters. 
\citet{2022arXiv221205984M} showed that at redshifts
$z \approx 2$ galaxies in high-density environments are more massive than galaxies
in low-density environments.
\citet{2020A&A...641A.172E} found that in the supercluster
A2142 even poor groups are probably merging. In contrast,
in low global luminosity--density environments 
there are no rich groups and clusters, meaning that even close to the filament axis
the local density enhancements are not high enough and groups are too
far apart to merge and form richer groups and clusters. 


Single galaxies and low-luminosity poor groups 
can be found everywhere in the cosmic web, from the lowest global
density  regions to the high-density cores of superclusters. 
Even almost one-half of single galaxies lie
far from filaments, and three-quarters of them lie at low global luminosity--density ($D8 \leq 2.38$).
This agrees  with the recent finding by \citet{2023arXiv230414387J} who showed 
using simulations that voids are mostly populated by 
haloes with mass below  $M \approx 10^{12} h^{-1} M_{\sun}$,
approximately the mass of the lowest luminosity groups (LL2.515).
\citet{2020MNRAS.493.1818D}  found that the poorest groups (pairs and triplets of galaxies)
reside in the lower density environments than  groups with four  galaxies or more.


\citet{2022A&A...668A..69E} mentioned that single galaxies 
in the  \citet{2014A&A...566A...1T} catalogue may also be outer members of rich clusters 
in which their  closest neighbour galaxies are faint, and therefore single galaxies are not connected with
other group members by FoF algorithm. This assumption was supported by the finding that in their
properties (e.g. the distributions of stellar masses and star formation properties)
single galaxies and satellites of high-luminosity groups are similar
\citep[see][for details]{2022A&A...668A..69E}. 
In order to be outer members of clusters, single galaxies have
to be located in their close neighbourhood. While a detailed analysis of this problem merits separate
study, we can give   some estimates here. Rich clusters lie in the global luminosity--density 
field with $D8 \geq 2.38$ and $D_{fil} \leq 2.5$~\Mpc. Our calculations show that approximately
$17$~\% of single galaxies lie in such environments, and only $3$~\% of single galaxies lie
in superclusters near the filament axis. Therefore, the comparison of the environments of rich clusters
and single galaxies shows that a large majority of single galaxies lie too far from 
clusters to be cluster members, and, most likely, they are the brightest galaxies
of faint groups. The similarity of the  properties of single galaxies and the BGGs of faintest groups
supports this conclusion.

We made a similar test for the lowest luminosity groups (LL2.5).
This test showed that $25$~\% of such groups lie in regions
with $D8 \geq 2.38$ and $D_{fil} \leq 2.5$~\Mpc, but only $5$~\%  of these lie
in superclusters with $D8 \geq 5$ and $D_{fil} \leq 2.5$~\Mpc. Therefore, as in the case of single galaxies,
a majority of the lowest luminosity groups lie too far from high-luminosity groups 
for infall.  

In addition, our test showed that some single galaxies may be 
misclassified  pair members (due to fibre collisions), and that 
galaxy pairs
may actually belong to triplets. We analyse pairs and triplets of galaxies together as
groups of the lowest luminosity, and thus, on average, the results do not change
because of this. We also found that the trends with environment, and the properties of 
single galaxies and the lowest luminosity groups are similar. Thus, we can say that
fibre collision effects do not change our results significantly.

Even if single galaxies and the lowest luminosity groups in our study lie far from
filaments, this does not mean that they form a random population in voids.
Although \citet{2022A&A...661A.115G} found some haloes not related to filaments,
observations and simulations suggest that galaxies with luminosity $M_r \leq -19.5$
(the luminosity limit in our study) form hierarchical filamentary network in voids, and
fainter galaxies are located in the same filamentary
structures as brighter ones \citep{Lindner:1995ui, 2023MNRAS.523.4693E}.

\subsection{Groups with different BGGs in the cosmic web. }
\label{sect:grbgg} 
 
In addition to differences in luminosity, richness, and 
environment, we found that the BGGs of  groups and clusters
are different. Clusters 
have mostly quenched BGGs (only $8$~\% of them have RSF BGGs, 
and 1--5 \%  are with BSF BGGs) with the highest 
stellar masses and luminosity and the lowest star formation rates among BGGs.
At the same time, more than one-half of poor groups have star-forming BGGs.

In clusters with a quenched BGG the BGG is located
closer to the group centre than in clusters with a star-forming BGG. This is an indicator that
clusters are also dynamically old, but galaxies in poor 
low-luminosity groups are still in the segregation stage.
As a support to this, \citet{2022A&A...668A..69E} found that in most groups with star-forming BGGs 
at least one member galaxy is quenched. They suggested that perhaps such groups are still forming,
and it has not yet been established which galaxy  will be central during future evolution.
In addition, \citet{2012A&A...540A.123E} found 
that in multimodal groups the BGG may lie far from the group centre. \citet{2012A&A...540A.123E} 
noted that typically
in such cases the BGG lies in one of the components of a group, but not in the main component.
This suggests that these groups are still forming.

Flattening of $\sigma^{\mathrm{*}}$ - $\sigma_{\mathrm{v}}$ at the high end in
Fig.~\ref{fig:sssvall} is in a good agreement with predictions  of  
the DIANOGA hydrodynamical zoom-in simulations \citep{2021MNRAS.507.5780M}. 
This flattening had already been observed in very early studies of
galaxy groups and clusters \citep{1976A&A....53...35E}.
This may be related to the growth of rich clusters. 
The stellar velocity dispersion $\sigma^{\mathrm{*}}$ of BGGs of some
clusters have values of the same order as  has the high end of velocity dispersions
$\sigma_{\mathrm{v}}$ of poor groups. This can be interpreted as a signature that 
the central part of a cluster
hosting present-day BGG started to form first. 
The BGGs in such clusters
were formed at early stages of group or cluster formation.
Then the group 
grows by the merging and infall of surrounding  galaxies and  groups that increase  overall 
velocity dispersion of a group, 
but do not affect the BGG strongly
\citep[see also discussion in][]{1998ApJ...502..141D,2000ASPC..209..360E, 2023A&A...676A.127D}.

One interesting result of our study is that the percentage of RSF BGGs among poor groups and single galaxies is very
stable, approximately  $17$\% in all environments. Only far from the filament axis ($D_{fil} \geq 2.5$~Mpc)
is the percentage of RSF BGGs   higher, reaching almost $25$~\% of all BGGs of
the lowest luminosity groups  and single galaxies
in these environments. 
The percentage of RSF BGGs among rich groups and clusters is also stable in all
environments, but, in contrast to poor groups, it
is much lower, approximately $8$~\%.
RSF galaxies are mostly of late type;  they  belong to the high-mass end of late-type
galaxies 
\citep[][]{2010MNRAS.405..783M,2011ApJ...736...51E, 2014MNRAS.440..889S, 2014A&A...562A..87E, 2018A&A...620A.149E}.
\citet{2014MNRAS.440..889S} noted that such galaxies are mostly central galaxies in
haloes with mass higher than  $M_{halo} \approx 10^{12} h^{-1} M_{\sun}$. In our sample this
mass approximately corresponds to the mass limit of LL2.515 groups. 
\citet{2018A&A...620A.149E} found that in groups of the supercluster SCl~A2142 RSF galaxies seem
to avoid central parts of clusters. Their sample of groups was small, and did not
exclude the possibility that these galaxies are the brightest galaxies in groups.
\citet{2014MNRAS.440..889S} and \citet{2022JCAP...03..024S}  related the phenomenon of red spirals with gas content of galaxies:
spiral galaxies with low cold gas fractions may be more easily quenched
and become red.

Another interesting finding is that the spread of luminosities of poor groups with $N_{gal} = 5$ with BSF BGGs,
both in superclusters and in voids is smaller than the spread of luminosities among groups
with red BGGs, and also smaller than among single galaxies. 
This is also seen in Fig.~\ref{fig:singlermagd8dcum}, where the spread of parameters of single galaxies is
larger than that of BGGs of low-luminosity groups. There are single galaxies with  higher stellar 
masses and luminosities than BGGs of low-luminosity groups. We may speculate that this has following
explanation. Galaxies, groups, and clusters form in dark matter
haloes of different mass:  the richer the system, the higher the  halo mass  needed for its formation.
Single galaxies (the brightest galaxies of faint groups) form in haloes
where the halo mass is not high enough for the formation of more than one luminous galaxy,
but, in   turn, the brightest galaxy in such a halo may be more massive and luminous than any
individual galaxy in a higher mass dark matter halo where more than one luminous galaxy can form.

\subsection{Processes at work to shape the properties of groups and their BGGs.}
\label{sect:proc}

\begin{figure}
\centering
\resizebox{0.44\textwidth}{!}{\includegraphics[angle=0]{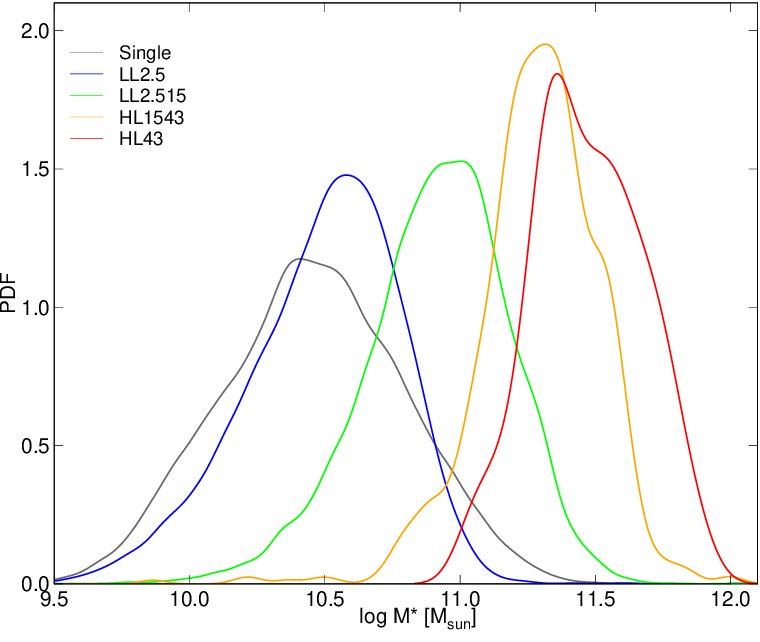}}
\caption{Distribution of stellar masses, $\log M^{\star}$,
for the BGGs in groups of various luminosity, and for single galaxies.
}
\label{fig:smdgrs}
\end{figure}
 
We showed that the properties of BGGs are different in groups of different luminosity.
In particular, the stellar masses of the BGGs increase from single galaxies to the BGGs 
of the richest clusters. To show this better we plot the distribution of stellar masses
$\log M^{\star}$ of BGGs and single galaxies in Fig. \ref{fig:smdgrs}, and provide median 
values of stellar masses in Table~\ref{tab:smassgrs}.
This
 suggests that processes that shape the properties of the BGGs are different (or have different
 timescales) in groups and clusters. 


\begin{table}[ht]
\centering
\caption{Median values of the stellar masses of BGGs.}
\begin{tabular}{rr} 
\hline\hline  
(1)&(2)\\      
\hline 
ID &   $\log M^*$  \\
\hline                                                    
HL43       & 11.46 \\ 
HL1543     & 11.30 \\ 
\hline                  
LL2.515    & 10.93 \\ 
LL2.5      & 10.52 \\ 
\hline                      
Single     & 10.46 \\ 
\hline                                        
\label{tab:smassgrs}  
\end{tabular}\\
\tablefoot{                                                                                 
Columns are as follows:
(1): ID;
(2): Median value of the stellar mass of BGGs $\log M^{\star}$.
}
\end{table}

Galaxy groups and clusters grow by infall  of individual galaxies and groups, 
and by the merging of groups and clusters \citep{2009ApJ...690.1292B,
2009MNRAS.400..937M, 2018MNRAS.477.4931H, 2020MNRAS.498.3852B}.
The growth of galaxies and their stellar mass, stochastic 
fluctuations of star formation rate, and finally star formation
quenching are governed by various processes, which are  divided 
into  `internal' and `external'  \citep{2006PASP..118..517B, 2020A&A...633A..70P,
Tacchella_2020, Patel_2023}.
Stellar masses of galaxies grow by mergers and gas infall. 
Internal processes, also called as mass quenching, depend first of all on the mass 
of the dark matter halo of a galaxy. 
Processes that blow out galactic gas include stellar winds, 
supernova explosions, and active galactic nucleus (AGN) feedback 
 \citep{2006MNRAS.372..265M, 2006MNRAS.365...11C,
2019MNRAS.485.3446H, Vulcani_2021}. These processes are more effective in massive galaxies 
 \citep{2020ApJ...889..156C} and at higher redshifts,
 although outflows of gas are also important   in low-mass galaxies.

External processes that eventually end the 
star formation in galaxies are caused by environmental quenching, which  depends on the local environment of galaxies within 
groups or clusters \citep{2019MNRAS.484.1702P, Vulcani_2021, 2021gcf2.confE..20W}. 
Such processes include the stripping of galactic gas by the ram 
pressure of the hot gas in a group or cluster 
\citep{1972ApJ...176....1G, 2019MNRAS.483.1042Y, Kolcu_2022, Herzog_2022, 2023arXiv230907037Z}, 
viscous stripping removing cold gas \citep{1982MNRAS.198.1007N}, 
starvation due to detachment from gas-feeding primordial filaments and cosmic web 
stripping 
\citep{2019OJAp....2E...7A, 2019A&A...621A.131M, Winkel_2021, Herzog_2022},
and harassment by high-speed mergers \citep{1996Natur.379..613M}. 
These processes depend on the density of the environment, 
on the orbital properties of galaxies, and also on galaxy mass, and they are more effective in 
less massive galaxies.

Single galaxies are stellar systems like our own Milky Way galaxy, surrounded by 
dwarf satellites in their dark matter halo, and without close neighbours in the same luminosity range.
The properties of single galaxies are determined by  processes in their dark matter haloes and intergalactic
medium around them. 
Single galaxies are fed by dwarf satellites, and by gas and dark matter from filaments in
immediate surroundings of a galaxy
\citep{Haywood:2016un, Bell:2017vy, Di-Matteo:2019wt,
2022A&A...666A.170Q}. Such events may disturb gas inflow from the surroundings and 
lead to a star formation quenching
in single galaxies  \citep{2016A&A...588A..79L, 2022ApJ...927..124M}.
Quenching of single galaxies may be a result of the cosmic web detachment (i.e.
detachment of primordial gaseous filaments),  which provided  galaxies with a fresh 
gas supply and supported star formation in them \citep{2019OJAp....2E...7A}.

In contrast to single galaxies, galaxies in low-luminosity groups have close neighbouring galaxies in the same
luminosity range. Thus, interactions with neighbouring galaxies
are not restricted to faint dwarf satellites, as in the  case of single
galaxies. This changes the character and frequency of mutual interactions.
Star-forming BGGs may be influenced by mergers or  interactions from neighbouring galaxies
\citep{2022MNRAS.515...22J}.
Most interactions with neighbouring galaxies are related
with the infall of neighbouring galaxies to the more massive (luminous)
galaxy. Impact angles of infalling galaxies depend on the richness of the systems
\citep{2021A&A...651A..56G, 2020MNRAS.493.4950S}.
In rich systems infall occurs mainly along the filaments surrounding clusters,
while in small systems the infall angles are more randomly distributed.
Only rarely is the impact 
directed towards the central galaxy, which leads to the merger of
the two galaxies and changes   the   properties of the central
galaxy. This may be the reason why 
the  trends with environment of the poorest groups and single galaxies
are similar. 

Star-forming BGGs obtain their gas from the surrounding medium or from stripped satellites
\citep{2019OJAp....2E...7A, 2022MNRAS.515...22J}.
If the gas supply stops, then the galaxy stops forming  stars
\citep[called   the cosmic web detachment in][]{2019OJAp....2E...7A}. The effectiveness of this
process depends on several factors, including the activity of the central supermassive black hole
\citep{2022MNRAS.515...22J}.
Very poor groups lie, on average, in a slightly higher density environment than
single galaxies. Therefore, the gas supply to increase the stellar mass
of galaxies, as well as the interactions with nearby galaxies 
may trigger star formation and then quenching in such groups
\citep[see also discussion about the properties of galaxies
in small groups by ][]{1982ApJ...255..382H, 2018MNRAS.481.2458D}.

\citet{2022JCAP...03..024S} found differences in both local and global environments 
of red and blue spirals, red spirals being located in higher density environments
than blue spirals. They assume that perhaps the spirals with
high cold gas content may have a blue colour, 
whereas the spirals with low cold gas content may have decreased
their star formation and become red.
Therefore, we can assume that in the faintest groups the local environment is 
more important than the whole group in shaping their BGGs.

Our analysis shows that the properties of LL2.515 groups and their BGGs
lie   between the  LL2.5 groups (and single galaxies) and the HL15 groups.
LL2.515 groups can grow by infall of single galaxies, and
by the mergers of galaxy pairs and triplets. Therefore,   their BGGs 
also have more supply to grow their stellar masses and quench their star formation.

Galaxies with their dark matter haloes are located in dark matter filaments. 
The enhanced evolution of groups and single galaxies 
in superclusters  may also be related to a larger amount of gas in filaments in global high-density regions near
clusters, as found in the simulations \citep{2021A&A...646A.156T}.
This again emphasizes the role of high-density environments in the evolution of
galaxies and groups.

\citet{2022MNRAS.515...22J} and  \citet{2023MNRAS.525.5677S} 
analysed the properties of central galaxies of group-sized haloes
from Romulus simulations, and 
the circumgalactic medium around  them. \citet{2023MNRAS.525.5677S} say that the presence
of cold gas and gaseous disks, rejuvenations, and the ongoing star
formation in BGGs of groups indicate that BGGs must be receiving inflow of gas from
their surroundings. Gas flows onto the central BGG via a filamentary
cooling flows and by infalling cold gas.
They describe  two pathways by which the gas that 
surrounds the BGGs cools. The first is via filamentary cooling inflows; another is via 
condensations forming from rapidly cooling density perturbations, which are mainly
seeded by orbiting substructures. 
In nearby groups, AGN feedback, stripping,  and both gas-rich and gas-poor mergers in the history of
BGGs are also important 
\citep{2017MNRAS.472.1482O, 2018A&A...618A.126O, 
2021IAUS..359..180K, 2022MNRAS.515.1104L}.
The importance of accretion in the evolution of BGGs in groups
is noted in \citet{2020A&A...635A..36G}.
These studies suggest that BGGs of groups evolve under the influence of 
various mechanisms, both internal and external \citep[][and references therein]{2022MNRAS.515.1104L}.

High-luminosity groups and clusters of galaxies are preferentially
located in global high-density regions in filaments or filament outskirts. 
All the richest and most luminous clusters reside in high-density cores of superclusters. 
In addition, the
spheres of dynamical attraction of rich clusters are much larger than those of groups
\citep{2021A&A...649A..51E}. 
A considerable fraction of interactions are due to infall
of neighbouring galaxies and groups along filaments, directed
to  clusters, especially in supercluster cores. This means that the growth of clusters
and the character and/or timescale of interactions between galaxies
changes in comparison with poor groups.
Galaxies fall into rich clusters along filaments, which also enhances
the quenching of galaxies in comparison with isotropic infall in poor groups
\citep{2020MNRAS.493.4950S}.
Quenched BGGs with very old stellar populations  formed their stars at least $10$~Gyrs ago
\citep[see][and references therein]{2022A&A...668A..69E}.
The Horizon Run 5 simulation suggests that in the densest regions of the cosmic web,
the birthplaces of the present-day rich galaxy clusters, 
galaxies may have started to form earlier than elsewhere
\citep{2022ApJ...937...15P, 2023arXiv230411911P}. 
This is also supported by observations of extremely high-redshift galaxies 
in forming protoclusters
\citep{Hashimoto_2023, 2023Natur.616..266L}.
\citet{2023MNRAS.523.3201D} proposed that massive very early galaxies may form by  
feedback-free starbursts,

At cluster scales, 
our results follow the trend observed in the  TNG and SIMBA simulations, 
where $78$~\% and $91$~\% of the BCGs, respectively, with low and high mass, 
are quenched
\citep{2021Univ....7..209O}. 
At group scales, the quenched fraction drops more in simulations 
(e.g. in Illustries, C-Eagle, Eagle-Ref, Romulus -C, TNG300) than in observations.
The differences between the simulations can be attributed 
to varying models for the interplay between AGN feedback and cooling procedures.
Using the  ROMULUS simulations, \citet{2022MNRAS.515...22J}  
  state  that efficient gas cooling from the CGM is an 
essential prerequisite for star formation in galaxy groups, 
while ram-pressure gas stripping from gas-rich satellites 
significantly supports suitable gas cooling flows. 
Galaxies in groups are often preprocessed (quenched) before they join the clusters, 
indicating a large-scale environmental dependence on galaxy properties 
\citep{2021Univ....7..209O, 2022A&A...668A..69E}.

The differences between the  processes in rich clusters and  poor groups were also highlighted
  in \citet{2022MNRAS.515...22J}. 
In rich clusters the BGGs lie in the cluster centres and their evolution is affected by mergers, galactic
cannibalism, dynamic friction, and other processes that lead to the formation of BGGs with the highest stellar mass and luminosity 
\citep[see e.g.][for a detailed discussion on the formation of BGGs of rich clusters]{2021MNRAS.507.5780M}.
In poor groups the velocities of galaxies are low, and the mergers of galaxies and 
strong tidal interactions are effective \citep{2022MNRAS.515...22J}.

\subsection{Summary}
\label{sect:summ} 
The results of our study of groups can be summarized as follows.

\begin{itemize}
\item[1)]
Groups can be divided into two main classes according to their luminosity, minimal
global luminosity--density at their location,  the star formation properties of their BBGs,
and distance from the filament axis. 
The threshold luminosity of groups
between the main classes is $L_{gr} = 15 \times10^{10} h^{-2} L_{\sun}$; the median mass  
is $M^{med}_{gr} \approx 23 \times10^{12} h^{-1} M_{\sun}$, being approximately ten times the local MW+M31 group mass.
We call these classes  clusters and poor groups.

Furthermore, clusters can be divided into two subclasses with luminosity limits
$L_{gr} \geq 43 \times10^{10} h^{-2} L_{\sun}$ and $15 \leq L_{gr} < 43 \times10^{10} h^{-2} L_{\sun}$, 
and poor groups  with luminosity limits 
$2.5 \leq L_{gr} < 15 \times10^{10} h^{-2} L_{\sun}$
and $L_{gr} \leq  2.5 \times10^{10} h^{-2} L_{\sun}$.
\item[2)]
Groups and clusters have BGGs with different star formation properties.
While $\approx 90$~\% of BGGs in clusters
are red and  quenched with no active star formation,  only approximately $40 - 60$~\% of the BGGs in 
poor groups and  of single galaxies  
are of this type. In poor groups and among single galaxies 
$\approx 17$~\% of BGGs are red star-forming galaxies (RSF),
and even $\approx 40$~\% of BGGs are blue star-forming (SF) galaxies.
These percentages are the same in a wide range of global luminosity--density.
\item[3)]
The location of high- and low-luminosity groups in the cosmic web is different.
Clusters
are absent in regions of low global
luminosity--density with  $D8 \leq 1.15$ ($D8 \leq 2.38$ for the richest clusters), and
they lie in filaments or filament outskirts,
with $D_{fil}  \leq 2.5$~\Mpc.
In contrast, poor groups and single galaxies reside everywhere in the cosmic web, from 
high-density cores of superclusters to voids. 
\item[4)]
Groups  of the same richness  in superclusters
are more luminous than in low-density regions between  superclusers. This difference is the largest
at richness $N_{\mathrm{gal}} = 10$, which is approximately the limiting richness
between groups and clusters.
Within a given richness class, groups with quenched BGGs have higher luminosity  than groups
with RSF BGGs, which in   turn are more luminous than groups with BSF BGGs.
\item[5)]
Quenched BGGs  are located closer to the group centre
than star-forming BGGs.
Rich groups and clusters follow  the relation between 
the stellar velocity dispersion of BGGs $\sigma^\star$ and
the group velocity dispersion $\sigma_{\mathrm{v}}$,  similar to that found in simulations.
\item[6)]  
Single galaxies are, on average,  fainter and less massive than BGGs of low-luminosity groups,
with a large spead in their parameter values. A total of 
18\% of single galaxies are RSF; this percentage is the same in various  global luminosity--density $ D8$ and
distance from filaments $D_{fil}$ intervals. 
Single galaxies are more luminous in superclusters than in low-density regions.
\end{itemize}

In summary, we obtained two main classes of galaxy systems: rich groups and clusters with mostly quenched BGGs,
and poor groups that may  also   have star-forming BGGs. The limiting luminosity between these classes is 
$L_{gr} =  15 \times10^{10} h^{-2} L_{\sun}$ and mass $M_{gr} \approx 23 \times10^{12} h^{-1} M_{\sun}$. 
Our results suggest that this is a threshold
luminosity between rich and poor galaxy systems, 
which also differ  in their BGG properties, dynamical properties, and location
in the cosmic web. Rich groups and clusters can only form in high global luminosity--density
regions near filaments,  and they are dynamically
older than poor groups.
The poorer the group and the farther it is located from filaments and rich structures (superclusters), 
the higher  the probability that  group has a star-forming BGG, either red or blue.
Single galaxies and  BGGs of low-luminosity groups seem to form a continuous sequence 
according to their properties, which supports the idea that single galaxies are the
brightest galaxies of faint groups. The birth and growth of BGGs in 
low-luminosity groups is different from the evolution of BGGs of high-luminosity groups.
Our study also emphasized the importance of different global environments where the formation
and growth of groups is different. Even groups of the same richness are more 
luminous in  global
higher density environments than in low-density environments.
Thus our study also highlights the role of high-density environments
(deep minima of the gravitation potential field) in the formation and evolution of
groups and clusters and their BGGs. 

To better understand the properties, environments, and possible evolution of groups, especially poor groups,
further studies are needed, which also  include   fainter galaxies than in this study.
Many aspects of the present study provide ideas for   future works. Among these is to 
use the data on fainter galaxies, groups, and filaments than in the present study,
for example, from the J-PAS survey \citep{2014arXiv1403.5237B} and especially the forthcoming
4MOST survey \citep{2019Msngr.175....3D, 2019Msngr.175...46D, 2023Msngr.190...46T}, and data from multiwavelength studies. 
It would be  interesting to analyse how the relation between the stellar velocity dispersion of BGGs and the 
velocity dispersion of groups and clusters is related to their substructure and connectivity.

\begin{acknowledgements}
We thank the referee for valuable comments and suggestions which helped us to
improve the paper. 
We thank Mirt Gramann for useful discussions.
We are pleased to thank the SDSS Team for the publicly available data
releases.  Funding for the Sloan Digital Sky Survey (SDSS) and SDSS-II has been
provided by the Alfred P. Sloan Foundation, the Participating Institutions,
the National Science Foundation, the U.S.  Department of Energy, the
National Aeronautics and Space Administration, the Japanese Monbukagakusho,
and the Max Planck Society, and the Higher Education Funding Council for
England.  The SDSS website is \texttt{http://www.sdss.org/}.
The SDSS is managed by the Astrophysical Research Consortium (ARC) for the
Participating Institutions.  The Participating Institutions are the American
Museum of Natural History, Astrophysical Institute Potsdam, University of
Basel, University of Cambridge, Case Western Reserve University, The
University of Chicago, Drexel University, Fermilab, the Institute for
Advanced Study, the Japan Participation Group, The Johns Hopkins University,
the Joint Institute for Nuclear Astrophysics, the Kavli Institute for
Particle Astrophysics and Cosmology, the Korean Scientist Group, the Chinese
Academy of Sciences (LAMOST), Los Alamos National Laboratory, the
Max-Planck-Institute for Astronomy (MPIA), the Max-Planck-Institute for
Astrophysics (MPA), New Mexico State University, Ohio State University,
University of Pittsburgh, University of Portsmouth, Princeton University,
the United States Naval Observatory, and the University of Washington.

The present study was supported by the ETAG projects 
PRG1006 and PSG700.
This work has also been supported by
ICRAnet through a professorship for Jaan Einasto. Suvi Korhonen acknowledges support by
the Vilho, Yrjö and Kalle Väisälä Foundation.
We applied in this study R statistical environment 
\citep{ig96}.

\end{acknowledgements}

\bibliographystyle{aa}
\bibliography{gr}

\end{document}